# Limits of stability for compounds of pentavalent praseodymium

Piotr G. Szkudlarek[1], Paweł Szarek[1,2]*, Wojciech Grochala[1]*

[1] Center of New Technologies, University of Warsaw, Zwirki i Wigury 93, 02-089 Warsaw, Poland
[2] Navi-Chem, Włodarzewska 83/120, 02-393 Warsaw, Poland

*This work is dedicated to Prof. Lester Andrews at his 80[th] birthday*

## Abstract

Eleven possible candidates for compounds of pentavalent praseodymium were investigated with relativistic density functional theory using the B3LYP functional with ZORA scalar relativistic correction and including spin-orbit coupling effects. Two of those candidates had previously been synthesized and another two of them were previously theoretically predicted. Three new species were proposed here as possible candidates for the compounds of Pr in its fifth oxidation state: $PrF_4^+$, $PrO_2F_2^-$ and $PrOF_2^+$. The main technical obstacle in synthesizing these ions is their high reactivity due to large electron affinity; decomposition via bimolecular reaction pathways and low energy of excitation to the triplet state constitute other stability limiting factors.

## Introduction

A concept of an oxidation state (OS) of chemical elements has been introduced to facilitate categorization of a multitude of chemical connections which are known for most elements. Since oxidation states are integer numbers (usually from −4 to +8, with some exceptions beyond that range which expand it to (−5, +9) [1–3]), categorization of chemical connections based on OSs greatly simplifies teaching of chemistry. In physics this approach is known simply as "ionic model". One usual procedure to assign an oxidation state in a connection between metallic and nonmetallic elements is to transfer appropriate number of electrons from a metal to nonmetallic ligands, so that Lewis's octet (or doublet, for H and He) rule is satisfied. The metal center is then left with a given number of holes, which is identical to the formal positive oxidation state of this central element. Such way of counting may easily be extended to connections between metallic and semi-metallic (or noble metal) elements, via so called Zintl-Klemm rules.[4,5] It has been shown that these simple concepts work beautifully even for anionic oxidation states of late TM elements.[6]

The concept of an OS abstracts from a true electron density (or charge) which could be assigned to each atomic center. However, its application permits to conduct comparative analyses of compounds of the same element (e.g. Ag, AgF, $AgF_2$, $AgF_3$, etc.) or of different elements (e.g. in the isoelectronic series: KF, $CaF_2$, $ScF_3$, $TiF_4$, etc.). Beautiful structural analogies emerge, which permit even making links between organic and inorganic chemistry (e.g. hexagonal structure of $Mg^{2+}(B^-)_2$ *vs.* that of graphite,[7] with $B^-$ and $C^0$ being isoelectronic; between benzene and its inorganic analogue, borazine; or between $H_3B-NH_3$ and ethane, etc.).

One early observation of historical and immense practical importance is that species bearing closed-shell cations are often very stable, thermodynamic sinks. This is exemplified by numerous chemical stoichiometries, and constitutes an extension of the 'doublet or octet' rule towards the 'decet' (for species with closed *d* shell) or 'tetradecet' (for lanthanides or actinides). The associated numbers of 2, 8, 10 and 14 stand, respectively, for closing of the *s*, *s+p*, *d* and *f* electronic subshells. Another rule of a thumb related to OSs is that stability (thermodynamic and thermal) usually decreases in the series of species featuring identical ligands and a core made of isoelectronic positively charged metal centers. This is beautifully exemplified by the following series:

$K_2O$, $CaO$, $Sc_2O_3$, $TiO_2$, $V_2O_5$, $CrO_3$, $Mn_2O_7$, $FeO_4$            (series I).

The series covers oxides of neighbouring *s* and *d* block-elements with formal oxidation state increasing from +1 to +8, but always adopting electronic configuration of a noble gas, Ar. Thermal and thermodynamic stability with respect to decomposition into an element (or a lower oxide) and oxygen strongly decreases in this series, with $K_2O$ being so stable that it may be melted at 740 °C without traces of decomposition, $TiO_2$ already being prone to release some $O_2$ from its surface when placed in ultra-high vacuum, $V_2O_5$ being in equilibrium with lower oxide and oxygen at sufficiently low temperatures to provide for its use as O-transfer catalyst, $CrO_3$ being unstable towards $Cr_2O_3$ formation in inert gas atmosphere, and $Mn_2O_7$ decomposing spontaneously in the explosive manner. No wonder that the last member of the series, $FeO_4$, has never been prepared, and it must be unstable even at very low temperatures.[8]

The trend of this kind holds nicely for many *s*, *p*, and *d* block elements as well as actinides, which to some extent resemble heavy *d* elements (Pa ≅ Ta, U ≅ W, Np ≅ Re, etc.). However, lanthanides most often escape rationalization based on isoelectronic principle, since their *f* electrons have semi-core nature. The properties of isoelectronic lanthanide cores are extremely different in any series and the cost of removal of a large a number of electrons is too large to take place in chemical compounds. In consequence, most lanthanides adopt the third (+3) oxidation state in their chemical connections. Still, remnants of the isoelectronic principle are seen in the following series which run within the *f*-block or across the *s/f* and *f/d* block boundaries:

$Rb^+$, $Ba^{2+}$, $La^{3+}$, **$Ce^{4+}$**         (series II),

**$Eu^{2+}$**, $Gd^{3+}$, **$Tb^{4+}$**         (series III) and

**$Yb^{2+}$**, $Lu^{3+}$, $Hf^{4+}$, etc.       (series IV).

The bold-font members, tetravalent Ce and Tb, as well as divalent Eu and Yb constitute the 'classical' non-trivalent Ln cations which occur in many chemical compounds of these elements. We notice that there are some others, too (e.g. $Sm^{2+}$, $Tm^{2+}$ etc.), but they do not fulfill the 'closed- or half-closed shell principle'.

Given that the isoelectronic principle for *d* block elements have been successfully extended to closed-shell $Ir^{9+}$,[9] and there is a good prospect to further extend it towards $Pt^{10+}$ (thus, OSs would span from −5 to +10), one is immensely tempted to try the same idea in the Ln series. An excursion towards the pentavalent state obviously seems to have the largest chances for success in series II and III, i.e. $Pr^{5+}$ and $Dy^{5+}$, respectively. While $Dy^{5+}$ has been shown to be too strong an oxidizer to bind to any ligands, the case of somewhat larger closed-shell cation, $Pr^{5+}$, turns out to be more promising. The pioneering 2015 study by Vent-Schmidt and Riedel has indicated the possibility of preparation of **$PrF_5$** (Figure 1k,l).[10] The $C_{4v}$ isomer was found to be by 5.1 kJ/mol more stable from the $D_{3h}$ one at the CCSD(T) level. Vibrational frequencies were calculated for both forms, and uni- and bimolecular channels for decomposition were studied. It was predicted that elimination of a F atom or $F_2$ molecule are energy-uphill, while $F_2$ removal via a bimolecular process is strongly exothermic. This could be the reason why $PrF_5$ has not been observed in experiment even at 20 K.

Subsequent experimental 2016 study has indicated the formation of **$PrO_2^+$** monocation (Figure 1f) and its adducts with Ar or $O_2$ molecule(s) (in the gas phase and in noble gas matrices).[11] Theoretical calculations have confirmed unequivocally the pentavalent nature of Pr in these species. Subsequent studies from the same group showed that it is possible to prepare **PrNO** molecule (Figure 1b) containing $Pr^{5+}$, which is quite prone to electron attachment leading to a monoanion.[12] Chemistry of the $PrO_2^+$ cation has also been extended to its connections with oxoanions such as $NO_3^-$ .[13] To understand trends in stability and chemical bonding, additional theoretical study was performed in 2020.[14] Its results have suggested that besides the already known species, **PrSO⁺**, **PrOF₃** and **PrOCl₃** could constitute viable synthetic targets. Given the fact that $Pr^{5+}$ in $PrF_5$ may oxidize $F^-$ anion to F radical with only small energy demand, renders the theoretical results for **PrSO⁺** and **PrOCl₃** somewhat dubious. These two species contain row III nonmetals, which loose electrons from their anions quite easily as compared to homologous row II elements (O, F). It may be presumed that careful studies of unimolecular or bimolecular decomposition channels for the said two species would clearly indicate their substantial instability.

Summarizing the results obtained during the last seven years we notice that two simple triatomic species of pentavalent praseodymium with ligands based on the most electronegative row II elements were successfully prepared, **PrO₂⁺** and **PrNO**. Two more were hypothetized (**PrF₅** and **PrOF₃**) but their preparation was not successful so far. The purpose of this theoretical study is to analyze properties of several isoelectronic molecular species of $Pr^{5+}$ including a few omitted from the previous studies, under one (quantum chemical methodology) roof, and to understand the limits and trends of stability of $Pr^{5+}$ compounds.

**Methodology**
Geometry optimization was performed using density functional theory (DFT) and the B3LYP[15–18] functional as implemented in the Amsterdam Density Functional (ADF) program ver. 2019.102 [19,20]. Relativistic effects were taken into account using scalar relativistic and spin-orbit coupling (SOC) effects using noncollinear approach in the zero-order regular approximation (ZORA)[21–23]. We also present results free from SOC corrections. In most cases default convergence criteria were used, however in some cases it was necessary to sharpen them slightly, especially for harmonic frequencies calculations. In these situations, criterium of energy change was set to $10^{-5}$ Hartree and the gradient criterium was set to $10^{-5}$ Hartree/Angstrom. Slater-type triple-ζ-plus two polarization functions were used.[24] Calculations were performed with no frozen core approximation. Orbital composition analysis, as well as visualization of spin densities was performed using the "Levels" module of ADFview. Orbitals visualization was performed in VESTA.[25] The values of contour of isovalues was 0.05. Visualization of the structures and vibration analysis was performed with freeware version of Chemcraft.[26] The dynamic stability of optimized structures was verified by performing numerical harmonic frequencies calculations.

**Results and discussion**
*1. Species studied.*
Here, in the spirt of our earlier studies for Ir- and Pt molecules, we have considered the following eleven species: the previously studied **PrO₂⁺**, **PrOF₃**, **PrNO**, and **PrF₅**, and seven novel ones: **PrO₂F₂⁻** anion, **PrOF₂⁺**, and **PrF₄⁺** cations, **PrN₂⁻** anion, as well as **PrNF₂** and related **PrNF₃⁻** anion and **PrNF⁺** cation. All of them feature ligands based on the most electronegative row II elements (F, O, N) and are either neutral (thus fulfilling the working definition of a chemical compound, which could possibly be 'bottled') or bear a −1 or +1 charge (and as such could be prepared either only in a very low concentration, either in conditions of a mass spectroscopy chamber, or inert gas matrices). We have deliberately avoided computing dications of any sort (which are usually prone to bond breaking via a redox process, or to Coulombic explosion) or dianions (most small molecular dianions tend to eliminate electron to vacuum). $PrF_5$ has five ligands, most of species considered here have four, **PrNF₂** and **PrOF₂⁺** - three, while **PrNF⁺**, **PrNO** and **PrN₂⁻** have two.

Writing their formulas, we have adopted a convention that less electronegative ligand is always listed before the more electronegative one (oxo-fluoride, nitride-oxide, nitride-fluoride, etc.). All stoichiometries are collected in Table 1.

| No ligands / charge | +1 | 0 | −1 |
|---|---|---|---|
| 2 | $PrNF^+$ <br> $PrO_2^{+}$*** [11] | $PrNO$** [12] | $PrN_2^-$ |
| 3 | $PrOF_2^+$ | $PrNF_2$ | [$PrNOF^-$] |
| 4 | $PrF_4^+$ | $PrOF_3$* [14] | $PrNF_3^-$ <br> $PrO_2F_2^-$ |
| 5 | --- | $PrF_5$* [10] | [$PrOF_4^-$] |
| 6 | --- | --- | [$PrF_6^-$] |

* previously hypothesized ** previously prepared in experiment

## 2. Molecular geometry.

As $Pr^{5+}$ is a closed shell species, and all ligands considered formally adopt octet configuration, one expects that species containing genuine $Pr^{5+}$ will take a singlet ground state. Molecular geometries for singlet states optimized at B3LYP/TZ2P level of theory (without SO correction) for all eleven species are shown in Figure 1 (a-l) together with corresponding bond lengths. Note, in section 4 we will discuss, inter alia, whether the said singlet states are indeed ground states of these systems. Please confront Supplementary Information for Cartesian coordinates of all molecules optimized (section S1).

A typical Pr–N bond length falls in the range of 1.65–1.76 Å; the Pr–O one is *ca.* 1.70–1.80 Å, while Pr–F one shows the largest span of 1.92–2.18 Å. The bond lengths increasing in the direction N < O < F clearly follow the expected trend for a triple bond to N, double one to O and single one to F. As the Pr=O bond length is the least variable in the series, one may adopt its average value of 1.75 Å, as well as use the tabularized Shannon-Prewitt radius of $O^{2-}$ anion (1.24 Å, the only available value is for tetrahedral environment rather than for terminal O atoms) to derive an ionic radius of $Pr^{5+}$ in the "ionic model"; one obtains the value of 0.54-0.56 Å in tetrahedral environment. The value in environment of two ligands only (i.e. in PrNO or $PrO_2^+$) is even smaller, down to 0.44-0.47 Å. While these values seem to be rather small (comparable, e.g. to $As^{5+}$ or $V^{5+}$ in tetrahedral environment, of 0.475 Å and 0.50 Å, respectively), they are not unrealistic.

To put these values in context, one needs to follow the $La^{3+}$, $Ce^{4+}$, $Pr^{5+}$ series in diverse ligand environments (Table 2). One may clearly see from Table 2 that there is a dramatic change of an "ionic radius" with respect to the metal coordination number (CN); e.g. for $La^{3+}$ the radius varies between 0.58 Å for mono-coordinated metal center in $LaO^+$ up to up to 1.38 Å for nona-coordinated one in hexagonal $La(OH)_3$ solid. This trend was visualized on the Figure 2. This is well-known, typical behavior of chemical bonds, which tend to be very short and quite covalent for small CNs and quite long and ionic for large CNs. To put it in yet another way, chemical potential of electron hungry $La^{3+}$ cation in the gas phase may be satisfied by bringing small share of electron density (hence, ionic bonding) from many ligands, or a large share of electron density from a single ligand at a short distance (hence, covalence for $LaO^+$). In each case, however, the chemical potential of both cation and anion become equal at a certain point along the bond in a stable species.

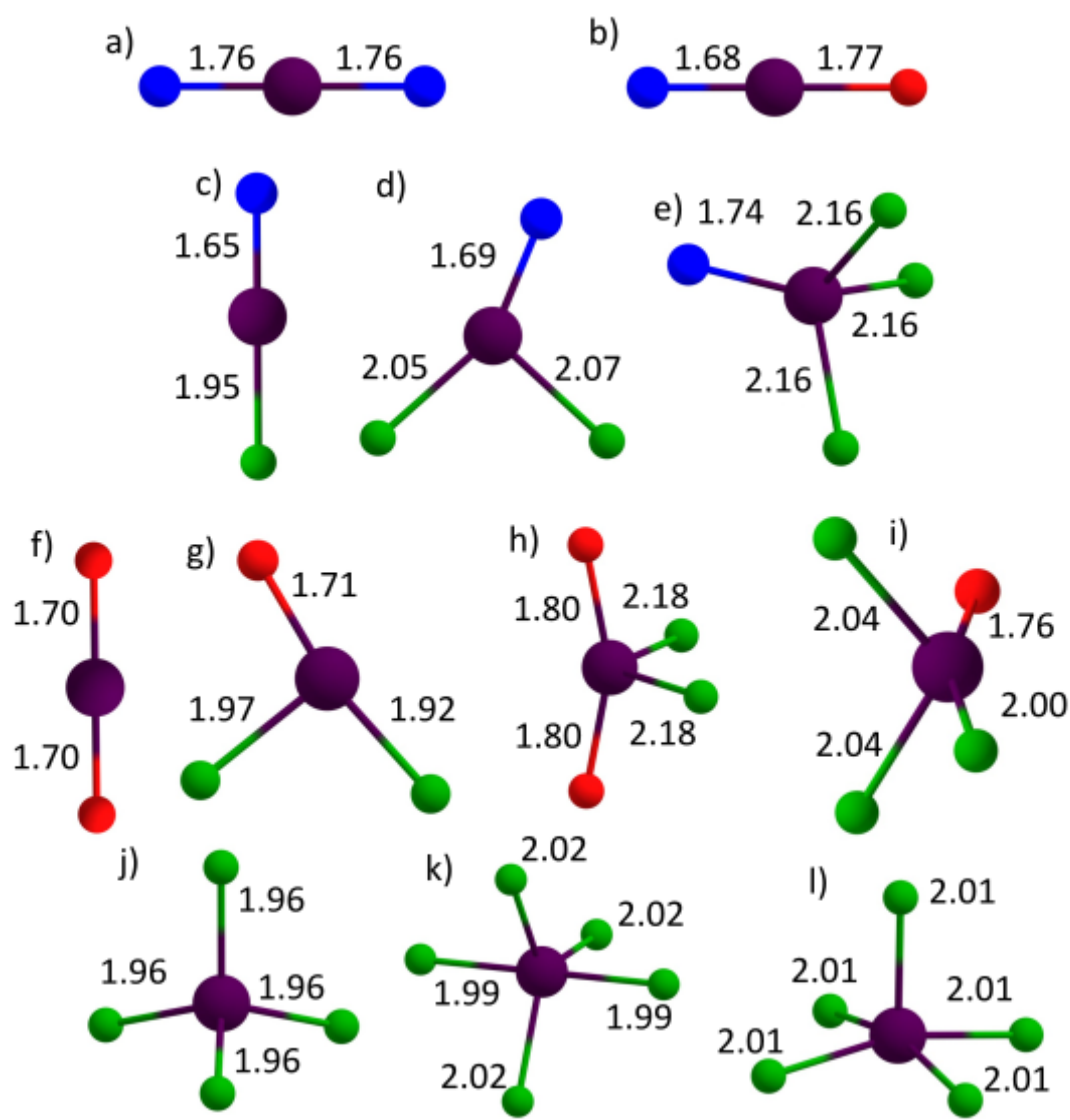

*Figure 1 Optimized geometries for investigated molecules: a) $PrN_2^-$, b) PrNO, c) $PrNF^+$, d) $PrNF_2$, e) $PrNF_3^-$, f) $PrO_2^+$, g) $PrOF_2^+$, h) $PrO_2F_2^-$, i) $PrOF_3$, j) $PrF_4^+$, k) $PrF_5$, bipyramid structure, l) $PrF_5$, square pyramid. All geometries are for singlet state without SOC correction.*

*Table 2 Ionic radii of $La^{3+}$, $Ce^{4+}$, $Pr^{5+}$ in diverse ligand environments. The radius of $O^{2-}$ of 1.24 Å was always assumed. CN stands for coordination number of a metal. S-P: Shannon-Prewitt values have been shown.*

| CN | $La^{3+}$ | $Ce^{4+}$ | $Pr^{5+}$ |
|---|---|---|---|
| 1 | 0.58 (LaO$^+$ cation, CCSD(T))[27] | --- | --- |
| 2 | 0.73 (LaO$_2^-$ anion, bent, $C_s$)[@] | 0.58 (CeO$_2$ molecule)[28] 0.60 (CeO$_2$ molecule, DFT qr)[29] | 0.44-0.47 (PrO$_2^+$, PrNO)[$] |
| 4 | 1.00 (La(OH)$_4^-$ anion, S$_4$)[@] | 0.85 (Ce(OH)$_4$ molecule)[28] | 0.54-0.56 (PrOF$_3$, PrO$_2$F$_2^-$)[$] |
| 6 | 1.17 (S-P) 1.19 (CuLaO$_2$ solid, hex)[30] | 1.00 (BaCeO$_3$ solid, cubic)[31] 1.01 (S-P) | --- |
| 8 | 1.30 (S-P) 1.35 (LaOF solid, hex)[32] | 1.11 (CeO$_2$ solid, cubic)[33] | --- |
| 9 | 1.35-1.38 (La(OH)$_3$ solid, hex)[34] | 1.29-1.32 (Ce(OH)$_3$ solid, hex)[35] | --- |

[@] This work; calculations at B3LYP/SDD level using scalar relativistic pseudopotentials and basis set.
[$] This work.

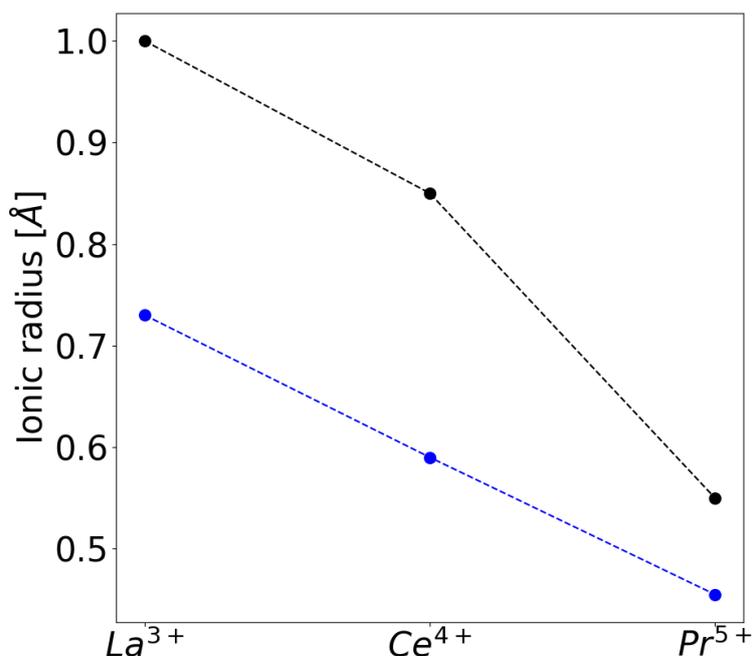

*Figure 2 Ionic radii for La$^{3+}$, Ce$^{4+}$ and Pr$^{5+}$ for coordination numbers 2 (black dots) and 4 (blue dots). If Table 2 reports a range, or more than one value, then arithmetic mean was plotted. Lines were added to guide an eye.*

The average values of 0.45 Å for bi- and 0.55 Å for tetra-coordinated Pr$^{5+}$ make perfect sense: (i) first, their ratio is ca. 1.22, not far from the respective ratio for Ce$^{4+}$ (1.44) and for La$^{3+}$ (1.37); (ii) secondly, their absolute values fit nicely the La$^{3+}$, Ce$^{4+}$, Pr$^{5+}$ series for identical CNs (e.g., 1.00 Å, 0.85 Å, and 0.55 Å, respectively, in tetrahedral environment). This contraction in isoelectronic series is large, and much more pronounced than the one in the series: Sc$^{3+}$, Ti$^{4+}$, V$^{5+}$ (0.74 Å$^{36}$, 0.56 Å, 0.50 Å, respectively), or in the series: Ga$^{3+}$, Ge$^{4+}$, As$^{5+}$ (0.61 Å, 0.53 Å, 0.475 Å, respectively). This has to do with the semi-core nature of the *f* states for the former series as contrasted with the valence nature of the d states in the latter. Nevertheless, in purely structural sense, Pr$^{5+}$, cation of a row VI element in tetrahedral environment may be thought of as a *ca.* 10-15% larger analogue of V$^{5+}$ or As$^{5+}$, cations of the row IV elements at their highest OS.

### 3. Vibrational frequencies.

The calculated harmonic normal mode frequencies have been tabularized in Table 3, and their ranges have been shown in Figure 3. No imaginary modes were detected at this level of theory which suggests that the studied singlets correspond to the local minima on the potential energy surface (PES). As expected, the highest frequencies for stretching of formally triple Pr≡N bonds (859–1068 cm$^{-1}$) exceed those for nominally double Pr=O ones (731–1017 cm$^{-1}$), and those in turn are usually larger than those typical of Pr–F single bonds (430–649 cm$^{-1}$).

The values of ligand-Pr-ligand bending and other deformation modes (for CN > 2) constitute a characteristic fingerprint of each species studied. The values derived here (*cf.* also SM, section S2) may be used to confirm the presence of the so-far hypothetical species in future experiments. This remark applies notably to these systems, for which singlet will prove to be the true ground state on the complex PES (*cf.* section 5).

The harmonic frequencies nicely follow the Badger rule (Figure 4)[37];*i.e.* the shorter the terminal chemical bond of some sort (say, Pr–F), the higher the corresponding stretching frequency. For example, the Pr-N bond length in the following species: **PrNF$^+$**, **PrNO**, **PrNF$_2$**, **PrNF$_3^-$** and **PrN$_2^-$**, equals to 1.65 Å, 1.68 Å, 1.69 Å, 1.74 Å, and 1.76 Å, respectively. Simultaneously, the respective highest-wavenumber Pr-N stretching frequencies follow the decreasing order: 1068, 1029, 954, 860, 859, and 907 cm$^{-1}$, with two exceptions from the monotonic trend.

*Figure 3 Vibrational frequencies for Pr-ligand stretching modes only for molecules in their singlet state. Color code indicates type of ligand: blue (N), red (O), green (F). Ranges of given mode types are shown as rectangles.*

*Figure 4 Illustration of fulfillment of the Badger's rule by Pr species studied here.*

*Table 3 Numerical harmonic normal mode frequencies calculated for singlet structures (frequencies for the corresponding spinorbit structures are given in parenthesis except for these cases when spinorbit structure resembles that of any higher spin state).*

| Molecule | $\bar{v}$ [cm$^{-1}$] | Assign. | Molecule | $\bar{v}$ [cm$^{-1}$] | Assign. | Molecule | $\bar{v}$ [cm$^{-1}$] | Assign. |
|---|---|---|---|---|---|---|---|---|
| PrN$_2^-$ | 129, 132 | δ $_{N-Pr-N}$ | PrOF$_2^+$ | 156 (156) | π $_{O-Pr-F}$ π $_{O-Pr-F}$ | PrF$_4^+$ | 43, 44, 44 (43, 44, 45) | δ $_{F-Pr-F}$ δ $_{F-Pr-F}$ |
| | 860 | ν $_{s\ N-Pr-N}$ | | 161 (160) | δ $_{F-Pr-F}$ δ $_{F-Pr-F}$ | | 63, 64 (64) | τ τ |
| | 907 | ν $_{as\ N-Pr-N}$ | | 180 (181) | δ $_{F-Pr-O}$ δ $_{F-Pr-O}$ | | 611 (611) | ν $_{as\ F-Pr-F}$ ν $_{as\ F-Pr-F}$ |
| PrNO | 158 | δ $_{O-Pr-N}$ | | 598 (599) | ν $_{as\ F-Pr-F}$ ν $_{as\ F-Pr-F}$ | | 622 (622) | ν $_{s\ F-Pr-F}$ ν $_{s\ F-Pr-F}$ |
| | 822 | ν $_{s\ O-Pr-N}$ | | 649 (649) | ν $_{s\ F-Pr-F}$ ν $_{s\ F-Pr-F}$ | PrF$_5$ | 54 (53) | δ $_{F-Pr-F}$ δ $_{F-Pr-F}$ |
| | 1029 | ν $_{as\ N-Pr-O}$ | | 901 (900) | ν $_{Pr-O}$ ν $_{Pr-O}$ | | 156 (156) | π $_{F-Pr-F}$ π $_{F-Pr-F}$ |
| PrNF$^+$ | 131, 133 | δ $_{F-Pr-N}$ | PrO$_2$F$_2^-$ | 82 | δ $_{F-Pr-F}$ | | 165 (165) | δ $_{F-Pr-F}$ δ $_{F-Pr-F}$ |
| | 619 | ν $_{s\ F-Pr-N}$ | | 178 | π $_{F-Pr-F}$ | | 179, 180 (180) | τ τ |
| | 1068 | ν $_{as\ F-Pr-N}$ | | 193 | δ $_{O-Pr-O}$ | | 492 (492) | ν $_{as\ F-Pr-F}$ |
| PrNF$_2$ | 116 | δ $_{N-Pr-F}$ | | 196 | τ | | 560 (560) | ν $_{as\ F-Pr-F}$ |
| | 119 | π $_{N-Pr-F}$ | | 237 | π $_{O-Pr-O}$ | | 586 (585) | ν $_{s\ F-Pr-F}$ |
| | 133 | δ $_{F-Pr-F}$ | | 430 | ν $_{s\ F-Pr-F}$ | | 586 (587) | ν $_{as\ F-Pr-F}$ |
| | 505 | ν $_{as\ F-Pr-F}$ | | 432 | ν $_{as\ F-Pr-F}$ | | | |
| | 557 | ν $_{s\ F-Pr-F}$ | | 731 | ν $_{s\ O-Pr-O}$ | | | |
| | 954 | ν $_{Pr-N}$ | | 806 | ν $_{as\ O-Pr-O}$ | | | |
| PrNF$_3^-$ | 100 | τ | PrOF$_3$ | 77 (77) | δ $_{F-Pr-F}$ δ $_{F-Pr-F}$ | | | |
| | 100 | δ $_{F-Pr-N}$ | | 143 (143) | δ $_{O-Pr-F}$ δ $_{O-Pr-F}$ | | | |
| | 121 | δ $_{N-Pr-F}$ | | 159 (158) | δ $_{F-Pr-F}$ δ $_{F-Pr-F}$ | | | |
| | 132 | π $_{F-Pr-F}$ | | 165 (166) | π $_{F-Pr-F}$ π $_{F-Pr-F}$ | | | |
| | 437 | ν $_{as\ F-Pr-F}$ | | 192 (193) | δ $_{O-Pr-F}$ δ $_{O-Pr-F}$ | | | |
| | 480 | ν $_{s\ F-Pr-F}$ | | 517 (518) | ν $_{as\ F-Pr-F}$ ν $_{as\ F-Pr-F}$ | | | |
| | 859 | ν $_{Pr-N}$ | | 546 (548) | ν $_{as\ F-Pr-F}$ ν $_{as\ F-Pr-F}$ | | | |
| PrO$_2^+$ | 202,206 | δ $_{O-Pr-O}$ | | 584 (584) | ν $_{s\ F-Pr-F}$ ν $_{s\ F-Pr-F}$ | | | |
| | 884 | ν $_{s\ O-Pr-O}$ | | 816 (816) | ν $_{Pr-O}$ ν $_{Pr-O}$ | | | |
| | 1017 | ν $_{as\ O-Pr-O}$ | | | | | | |

## 4. Bond order (BO), electronic population and molecular orbital (MO) analyses.

Proper description of chemical bonding to highly charged Ln core is not an obvious task, as the electrostatic and chemical potential changes dramatically at short distance around the cation. To get additional insight into chemical bonding in molecules studied we have performed two types of BO analyses, the G-J[38] and N-M[39]. These indexes were used in the past to rationalize electron density distribution for many different types of chemical bonds, with good success. The values of BO have been collected in S3 in **SM**. The excerpt is presented in Table 4. Here we will briefly analyze the values obtained.

*Table 4* Bond lengths in optimized singlet structures and two values of bond order according to G-J and N-M definitions.

| Formula | Bond lenght [Å] | $BO_{G-J}$ | $BO_{N-M}$ |
|---|---|---|---|
| Pr-F bonds | | | |
| $PrNF^+$ | 1.945 | 0.70 | 1.29 |
| $PrNF_2$ | 2.066 | 0.58 | 1.15 |
| $PrNF_2$ | 2.046 | 0.56 | 1.06 |
| $PrNF_3^-\ [x3]$ | 2.155 | 0.44 | 0.94 |
| $PrOF_2^+$ | 1.966 | 0.86 | 1.48 |
| $PrOF_2^+$ | 1.925 | 0.81 | 1.35 |
| $PrO_2F_2^-\ [x2]$ | 2.177 | 0.41 | 0.92 |
| $PrOF_3[x2]$ | 2.043 | 0.65 | 1.18 |
| $PrOF_3[x1]$ | 2.005 | 0.67 | 1.14 |
| $PrF_4^+[x4]$ | 1.958 | 0.98 | 1.47 |
| $PrF_5[x3]$ | 2.019 | 0.78 | 1.26 |
| $PrF_5[x2]$ | 1.994 | 0.82 | 1.28 |
| Pr-N bonds | | | |
| $PrN_2^-$ | 1.755 | 2.12 | 3.20 |
| $PrNO$ | 1.682 | 2.34 | 3.64 |
| $PrNF^+$ | 1.647 | 2.58 | 4.28 |
| $PrNF_2$ | 1.691 | 2.52 | 3.91 |
| $PrNF_3^-$ | 1.739 | 2.58 | 3.79 |
| Pr-O bonds | | | |
| $PrNO$ | 1.775 | 1.48 | 2.44 |
| $PrO_2^+$ | 1.696 | 1.75 | 2.88 |
| $PrOF_2^+$ | 1.707 | 1.94 | 3.02 |
| $PrO_2F_2^-$ | 1.802 | 1.52 | 2.42 |
| $PrOF_3$ | 1.755 | 1.85 | 2.80 |

The N-M scheme tends to somewhat overestimate the order of chemical bonding in the species studied. The $BO_{N-M}$ values for triple Pr-N bonds fall in the 3.20-4.28 range, those for the double Pr-O bonds in the 2.42-3.02 one, and the values for the single Pr-F bonds are 0.92-1.48. On the other hand, the G-J scheme yields more realistic values of $BO_{G-J}$: 2.12-2.58 for Pr-N, 1.48-1.94 for Pr-O and 0.41-0.98 for Pr-F. In each case the latter values do not exceed 3, 2 and 1, respectively. Yet, the uppermost values are not very far from the corresponding integers (particularly for bonds to O or F), thus pointing to substantial covalent character of the chemical bonding in these species.

This is independently confirmed through Mulliken and Hirschfeld population analyses (cf. SM, section S4), which suggest that a charge of +1.16-2.54 e (Mulliken) or +1.08-1.21 e (Hirschfeld) resides at Pr center in overall **neutral** species. The nitride, oxide and fluoride ligands bear Hirschfeld charges from −0.32 e to −0.58 e (N), from −0.56 e to +0.03 e (O), and from −0.47 e to +0.02 e (F), respectively. Such charge distribution also implies substantial covalence of chemical bonding and quasi-atomic nature of (formally anionic) ligands.

The same picture obtains from analysis of molecular orbitals (MOs) (cf. SM, section S5), which point out to very small (fractional) occupation of $f$ orbitals in Pr in all occupied orbitals, and substantial one in several unoccupied orbitals above Fermi level of these molecules. Binding energies of electrons sitting in HOMO are very large for neutral systems (from 6.0 to 11.5 eV, cf. SM, section S6) suggesting that it is not possible to additionally ionize these systems using chemical means. These values exceed the one computed for non-dipolar surface of $AgF_2$, which is known to be extremely strong oxidizer. The respective values for anionic species drop substantially, as expected (down to 2.5-2.9 eV). Now, ionization seems to be within reach, but electron would be removed from ligands rather than Pr center. $PrN_2^-$ is special by having a formally negative ionization energy which means that such species would spontaneously loose electron in vacuum.

The Pr-centered LUMO governs the electron acceptor properties and serve as approximate estimate of an electron affinity. Energies of LUMO for neutral compounds studied range from 4.1 eV (for $PrNF_2$) to 7.5 eV (for $PrF_5$). The latter value is already immense, but – as expected - it is surpassed by that for the $PrF_4^+$ cation (14.2 eV). This cationic species would steal electrons from every possible source in any realistic chemical environment.

The HOMO-LUMO gap is the largest for $PrO_2^+$ among all species studied here (4.4 eV), and the smallest for $PrN_2^-$ (2.4 eV). The electronic hardness (equal half of the said gap) governs kinetic stability of a system, both internal, as well as connected with external factors, and that is one essential component of reactivity in any real chemical environment. We will return to hardness in section 6. Here we will notice that the HOMO-→LUMO excitation (with the resulting hole-electron attraction), which must be of the ligand-to-metal charge-transfer character, will likely fall in near UV. The species studied will be colourless or yellow, depending whether the absorption edge would extend into the visible region of electromagnetic spectrum. Energies of the lowest singlet – singlet transition and the lowest singlet - triplet vertical transition were collected in the **Table 5**.

*Table 5 The lowest energies of the singlet - singlet and singlet - triplet vertical transitions for molecules studied.*

| Compound | Lowest singlet − singlet transition energy [eV] | Lowest singlet − triplet transition energy [eV] |
|---|---|---|
| $PrN_2^-$ | * | * |
| PrNO | 0.38 | -0.30 |
| $PrNF^+$ | NC | NC |
| $PrNF_2$ | 0.37 | -0.29 |
| $PrNF_3^-$ | 0.11 | -1.64 |
| $PrO_2^+$ | 1.02 | 0.27 |
| $PrOF_2^+$ | 1.33 | 0.77 |
| $PrO_2F_2^-$ | 1.14 | 0.59 |
| $PrF_3O$ | 1.46 | 0.93 |
| $PrF_4^+$ | 2.11 | 1.76 |
| $PrF_5$ | 2.33 | 1.87 |

* spontaneous emission of electron in vacuum; NC – not converged

*5.   Other multiplicities and calculations including Spin-Orbit (SO) coupling.*

So far, we have discussed solely the optimized singlet states of our systems. However, the inherent nature of systems bearing high oxidation states of a metal is such, that some of them will rather tend to undergo a spontaneous electron transfer from the ligand(s) to the metal core (an internal redox process). This happens in many systems of this type as a manifestation of "redox non-innocence" of even fluoride ligands. One natural consequence of that is that at least one chemical bond is weakened (or even broken), and spin density appears on both metal and ligand. This can be followed by additional processes, such as bond formation between ligands, etc. In every case this implies lowering of the formal OS of a metal. In the case of $Pr^{5+}$ singlet precursor, this would lead to a $Pr^{4+}$ system (triplet) or a very common $Pr^{3+}$ derivative (quintet).

In Table 6 we show the energies of the triplet and quintet states of the molecules studied (optimized at the same level as singlet) referred to the energy of the optimized singlet.

*Table 6* Energies of the lowest triplet and quintet states referred to energy of the optimized singlet.

| Formula | Multiplicity | ΔE [kcal/mol] | ΔH [kcal/mol] | ΔG [kcal/mol] |
|---------|--------------|---------------|---------------|---------------|
| $PrN_2^-$ | Triplet | -1.6 | -1.9 | -1.7 |
| | Quintet | 39.3 | 38.7 | 37.7 |
| $PrNO$ | Triplet | 4.3 | 4.0 | 3.8 |
| | Quintet | 33.5 | 32.3 | 30.0 |
| $PrNF^+$ | Triplet | -5.0 | -5.6 | -7.1 |
| | Quintet | 3.6 | 2.7 | 1.1 |
| $PrNF_2$ | Triplet | -4.5 | -4.8 | -5.6 |
| | Quintet | -5.0 | -5.8 | -7.0 |
| $PrNF_3^-$ | Triplet | -6.6 | -6.9 | -7.0 |
| | Quintet | -5.9 | -6.4 | -6.9 |
| $PrO_2^+$ | Triplet | 46.1 | 45.7 | 45.5 |
| | Quintet | 67.9 | 66.6 | 64.0 |
| $PrOF_2^+$ | Triplet | 8.3 | 7.9 | 7.3 |
| | Quintet | 15.3 | 14.5 | 11.4 |
| $PrO_2F_2^-$ | Triplet | 10.9 | 10.4 | 8.5 |
| | Quintet | 35.2 | 34.4 | 32.3 |
| $PrOF_3$ | Triplet | 3.9 | 3.6 | 3.2 |
| | Quintet | 21.5 | 20.7 | 18.9 |
| $PrF_4^+$ | Triplet | 14.6 | 14.4 | 15.2 |
| | Quintet | 41.6 | 41.1 | 42.7 |
| $PrF_5$ | Triplet | 5.3 | 4.4 | 4.7 |
| | Quintet | 60.0 | 59.3 | 57.2 |

It is clear from the Table 5 that for several systems singlet is <u>not</u> the ground state. This is the case for $PrN_2^-$, $PrNF_3^-$, $PrNF_2$, and $PrNF^+$. In these cases, either triplet or quintet has the lowest energy. In the case $PrNO$ the triplet state is only 4.3 kcal/mol higher than singlet. The internal redox process is immediately visible in spin densities on atoms (Table 7) and in substantial decrease of vibrational frequency of these bonds, which have been weakened by the excitation (cf. SM, section S2). The BO analysis also confirms this rather obvious scenario.

*Table 7* Spin densities of the ground state structures of three systems with the ground state being either triplet or quintet.

| Formula | Multiplicity | Pr | N | F |
|---------|--------------|-----|-----|-----|
| $PrNF_2$ | Quintet | 2.1 | 1.9 | 0 |
| $PrNF_3^-$ | Triplet | 1.4 | 0.6 | 0 |
| $PrNF^+$ | Triplet | 1.7 | 0.4 | 0 |
| Formula | Multiplicity | Pr | N | |
| $PrN_2^-$ | Triplet | 1.8 | 0.1 | |

In the remaining cases, singlet is clearly the ground state. Here, the lowest excited electronic configuration (in its optimized geometry) falls from 5.3 kcal/mol for $PrF_5$ to 46.1 kcal/mole for $PrO_2^+$. While the lower boundary of these values does not guarantee sufficient stability at ambient temperature conditions, the molecules in question should survive at helium temperatures without undergoing any destructive internal redox process. We notice that substantial stability of $PrO_2^+$ is in line with the fact that this cation was prepared in experiment.

The ultimate cross-check of stability of the singlet configuration was performed using calculations including the SO coupling. Here, we will very briefly comment the outcome.

First, the SO calculations have confirmed an inherently unstable nature of all species, for which singlet was detected to be not a ground state (see discussion above). In such cases, the SO optimized geometry resembles that for either triplet or quintet states, and the harmonic frequencies at the SO level also correlate well with those for states of high multiplicities. On the other hand, $PrF_5$ and $PrF_4^+$ were confirmed to adopt a geometry and harmonic frequency pattern which resembled that computed for the singlet state. The most interesting results were obtained for $PrO_2^+$, $PrOF_3$ and $PrO_2F_2^-$. In the first case, SO calculations predict the optimum geometry to correspond to that of a triplet state rather than singlet. The fact that $PrO_2^+$ has been observed in experiment may be explained partially by the deficiencies of theory, and partly by the fact that these species in fact exist in noble gas matrices, where extra source of stabilizing electron density in the form of argon atoms is available.

For $PrOF_3$, the butterfly minimum detected clearly has a singlet-like characteristics, by having a very short Pr-O bond (1.756 Å), with very high stretching frequency of 815 cm$^{-1}$. Thus, this is a genuine pentavalent Pr system.

The last important result is that the SO calculations predict a triplet-like geometry for $PrO_2F_2^-$ (vibrational characteristics also matches such assignment). Recall that the calculations without taking SO into account have suggested that this molecule has a singlet ground state, and an appreciable singlet-triplet energy gap.

## 6. Uni- and bimolecular decomposition channels as possible kinetic stability limiting factors.

We have studied diverse hypothetical unimolecular decomposition channels for all singlet states discussed in section 7. Moreover, we have computed thermodynamics of selected bimolecular events which were expected to be relatively exoenergetic. Here we show excerpt only for systems for which either a calculation excluding or the one involving SO coupling would suggest the presence of $Pr^{5+}$ in the ground state (Table 8). The relevant data for molecules with non-singlet ground state was presented in S7. Note that the SO calculations tend to more reliably reproduce the experimental values for several reference reactions (e.g., $F_2 \rightarrow 2$ F, etc.).

*Table 8* Calculated thermochemistry for singlet structures of different singlet praseodymium systems for selected uni- and bimolecular reaction channels.

| Equation | Calculations | ΔE [kcal/mol] | ΔH [kcal/mol] | ΔG [kcal/mol] |
|---|---|---|---|---|
| $PrNO \rightarrow PrN + {}^{3}O$ | Scalar | 163.3 | 162.1 | 154 |
| $PrNO \rightarrow PrN + {}^{3}O$ | Spinorbit | 133.9 | 133.2 | 125.1 |
| $2\,PrNO \rightarrow 2\,PrN + O_2$ | Scalar | 142.6 | 141.8 | 132.2 |
| $2\,PrNO \rightarrow 2\,PrN + O_2$ | Spinorbit | 141.2 | 141.3 | 131.8 |
| $PrOF_2^+ \rightarrow PrOF^+ + F^{\bullet}$ | Scalar | 83.4 | 82.9 | 75.4 |
| $PrOF_2^+ \rightarrow PrOF^+ + F^{\bullet}$ | Spinorbit | 46.4 | 45.8 | 38.2 |
| $PrOF_2^+ \rightarrow PrF_2^+ + {}^{3}O$ | Scalar | 55.3 | 54.2 | 46.1 |
| $PrOF_2^+ \rightarrow PrF_2^+ + {}^{3}O$ | Spinorbit | 26.6 | 25.5 | 17.3 |
| $2\,PrOF_2^+ \rightarrow 2\,PrF_2^+ + O_2$ | Scalar | -73.2 | -74 | -83.7 |
| $2\,PrOF_2^+ \rightarrow 2\,PrF_2^+ + O_2$ | Spinorbit | -73.5 | -74.2 | -83.9 |
| $2\,PrOF_2^+ \rightarrow 2\,PrOF^+ + F_2$ | Scalar | 58.9 | 58.5 | 49.8 |
| $2\,PrOF_2^+ \rightarrow 2\,PrOF^+ + F_2$ | Spinorbit | 50.9 | 50.4 | 41.5 |
| $PrO_2F_2^- \rightarrow PrF_2 + O_2^-$ | Scalar | 135.9 | 134.9 | 124.5 |
| $PrO_2F_2^- \rightarrow PrF_2 + O_2^-$ | Spinorbit | 124.6 | 124.1 | 115.3 |
| $PrO_2F_2^- \rightarrow PrF_2^- + O_2$ | Scalar | 119.8 | 119.4 | 109.1 |
| $PrO_2F_2^- \rightarrow PrF_2^- + O_2$ | Spinorbit | 109.7 | 109.2 | 100.5 |
| $PrO_2F_2^- \rightarrow PrOF + OF^-$ | Scalar | 139.6 | 138.4 | 127.3 |
| $PrO_2F_2^- \rightarrow PrOF + OF^-$ | Spinorbit | 134.8 | 133.7 | 123.6 |
| $PrO_2F_2^- \rightarrow PrOF^- + OF^{\bullet}$ | Scalar | 171.0 | 170.1 | 159.1 |
| $PrO_2F_2^- \rightarrow PrOF^- + OF^{\bullet}$ | Spinorbit | 160.5 | 160.1 | 150.7 |
| $PrO_2F_2^- \rightarrow PrO_2^- + F_2$ | Scalar | 198.4 | 197.7 | 187.7 |
| $PrO_2F_2^- \rightarrow PrO_2^- + F_2$ | Spinorbit | 188.1 | 187.9 | 179.5 |
| $PrO_2F_2^- \rightarrow PrO_2 + F_2^-$ | Scalar | 162.0 | 160.9 | 150.4 |
| $PrO_2F_2^- \rightarrow PrO_2 + F_2^-$ | Spinorbit | 116.0 | 115.5 | 106.9 |
| $PrO_2F_2^- \rightarrow PrF_2^+ + O_2^{2-}$ | Scalar | 458.6 | 457.1 | 446.7 |
| $PrO_2F_2^- \rightarrow PrF_2^+ + O_2^{2-}$ | Spinorbit | 494.3 | 493.3 | 484.5 |
| $PrO_2F_2^- \rightarrow PrO_2F^- + F^{\bullet}$ | Scalar | 131.6 | 130.8 | 121.4 |
| $PrO_2F_2^- \rightarrow PrO_2F^- + F^{\bullet}$ | Spinorbit | 88.0 | 87.7 | 79.8 |
| $PrO_2F_2^- \rightarrow PrOF_2^- + {}^{3}O$ | Scalar | 117.3 | 116.2 | 106.2 |
| $PrO_2F_2^- \rightarrow PrOF_2^- + {}^{3}O$ | Spinorbit | 78.7 | 78.0 | 69.5 |
| $PrO_2^+ \rightarrow Pr^+ + O_2$ | Scalar | 129.0 | 128.7 | 122.4 |
| $PrO_2^+ \rightarrow Pr^+ + O_2$ | Spinorbit | 114.9 | 115.2 | 109.1 |
| $PrO_2^+ \rightarrow Pr + O_2^+$ | Scalar | 285.9 | 286.3 | 280.1 |
| $PrO_2^+ \rightarrow Pr + O_2^+$ | Spinorbit | 381.1 | 380.7 | 374.4 |
| $PrO_2^+ \rightarrow PrO^+ + {}^{3}O$ | Scalar | 119.9 | 118.7 | 109.7 |
| $PrO_2^+ \rightarrow PrO^+ + {}^{3}O$ | Spinorbit | 67.3 | 66.5 | 57.7 |
| $2\,PrO_2^+ \rightarrow 2\,PrO^+ + O_2$ | Scalar | 55.9 | 54.9 | 43.6 |
| $2\,PrO_2^+ \rightarrow 2\,PrO^+ + O_2$ | Spinorbit | 7.9 | 7.8 | -3.0 |
| $2\,PrO_2F_2^- \rightarrow PrO_2F^- + PrOF_2^- + OF^{\bullet}$ | Scalar | 155.3 | 154.2 | 141.9 |
| $2\,PrO_2F_2^- \rightarrow PrO_2F^- + PrOF_2^- + OF^{\bullet}$ | Spinorbit | 134.2 | 134.1 | 124.8 |
| $2\,PrO_2F_2^- \rightarrow 2\,PrOF_2^- + O_2$ | Scalar | 50.8 | 49.9 | 36.5 |
| $2\,PrO_2F_2^- \rightarrow 2\,PrOF_2^- + O_2$ | Spinorbit | 30.8 | 30.9 | 20.6 |
| $2\,PrO_2F_2^- \rightarrow 2\,PrO_2F^- + F_2$ | Scalar | 155.3 | 154.2 | 141.9 |
| $2\,PrO_2F_2^- \rightarrow 2\,PrO_2F^- + F_2$ | Spinorbit | 134.2 | 134.1 | 124.8 |
| $2\,PrO_2F_2^- \rightarrow 2PrO_2F^- + PrOF_2 + OF^-$ | Scalar | 129.3 | 128 | 114.3 |
| $2\,PrO_2F_2^- \rightarrow 2PrO_2F^- + PrOF_2 + OF^-$ | Spinorbit | 113.8 | 112.9 | 101.8 |
| $PrOF_3 \rightarrow PrF_2 + OF^\bullet$ | Scalar | 126.8 | 125.8 | 115.6 |
| $PrOF_3 \rightarrow PrF_2 + OF^\bullet$ | Spinorbit | 124.8 | 123.8 | 113.8 |
| $PrOF_3 \rightarrow PrOF + F_2$ | Scalar | 125.5 | 124.7 | 114.8 |
| $PrOF_3 \rightarrow PrOF + F_2$ | Spinorbit | 125.6 | 124.8 | 115 |
| $PrOF_3 \rightarrow PrF_3 + {}^3O$ | Scalar | 61.2 | 60.1 | 51.4 |
| $PrOF_3 \rightarrow PrF_3 + {}^3O$ | Spinorbit | 33.7 | 32.6 | 24.3 |
| $PrOF_3 \rightarrow PrOF_2 + F^\bullet$ | Scalar | 91.6 | 90.8 | 81.4 |
| $PrOF_3 \rightarrow PrOF_2 + F^\bullet$ | Spinorbit | 58.5 | 57.7 | 48.1 |
| $2\,PrOF_3 \rightarrow 2\,PrF_3 + O_2$ | Scalar | -61.6 | -62.3 | -73.1 |
| $2\,PrOF_3 \rightarrow 2\,PrF_3 + O_2$ | Spinorbit | -59.4 | -60.0 | -70.0 |
| $PrF_4^+ \rightarrow PrF_2^+ + F_2$ | Scalar | 34.6 | 33.9 | 27.3 |
| $PrF_4^+ \rightarrow PrF_2^+ + F_2$ | Spinorbit | 35.0 | 34.3 | 27.3 |
| $PrF_4^+ \rightarrow PrF_3^+ + F^\bullet$ | Scalar | 62.8 | 62.1 | 55.3 |
| $PrF_4^+ \rightarrow PrF_3^+ + F^\bullet$ | Spinorbit | 14.9 | 14.3 | 9.0 |
| $2\,PrF_4^+ \rightarrow 2PrF_3^+ + F_2$ | Scalar | 17.6 | 17 | 9.7 |
| $2\,PrF_4^+ \rightarrow 2PrF_3^+ + F_2$ | Spinorbit | -12.1 | -12.6 | -16.9 |
| $PrF_5 \rightarrow PrF_3 + F_2$ | Scalar | 62.2 | 61.4 | 51 |
| $PrF_5 \rightarrow PrF_3 + F_2$ | Spinorbit | 63.8 | 62.9 | 52.3 |
| $PrF_5 \rightarrow PrF_4 + F^\bullet$ | Scalar | 56.1 | 55.4 | 45.8 |
| $PrF_5 \rightarrow PrF_4 + F^\bullet$ | Spinorbit | 23 | 22.3 | 12.4 |
| $2\,PrF_5 \rightarrow 2\,PrF_4 + F_2$ | Scalar | 4.3 | 3.6 | -9.4 |
| $2\,PrF_5 \rightarrow 2\,PrF_4 + F_2$ | Spinorbit | 4.2 | 3.4 | -10.1 |

For $PrF_5$ the unimolecular decomposition channel which is characterized by the smallest positive enthalpy is that of dissociating one F atom:

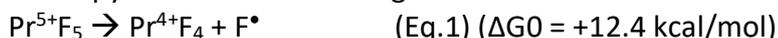
$Pr^{5+}F_5 \rightarrow Pr^{4+}F_4 + F^\bullet$       (Eq.1) ($\Delta G0 = +12.4$ kcal/mol)

A qualitatively different result was obtained by Vent-Schmidt and Riedel[10] (−2.6 kcal/mol) but their results did not take SOC into account. A homolytic bond breaking costs *ca.* 22 kcal/mol (Vent-Schmidt & Riedel[10]: 29.3 kcal/mol) and this indicates that Pr-F bond is rather weak (*ca.* 1 eV) as for a single bond. Note, the strongest single bond in neutral diatomics is that of $H_2$, some 4.5 eV. However, this process is characterized by even smaller $\Delta G^0$ of *ca.* 12 kcal/mol due to entropy factor, which certainly limits the thermal stability of $PrF_5$ molecule in isolation. Most importantly, the associated bimolecular decomposition channel leading to $F_2$ molecule:

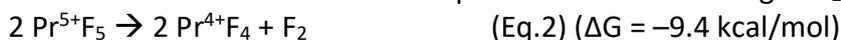
$2\,Pr^{5+}F_5 \rightarrow 2\,Pr^{4+}F_4 + F_2$       (Eq.2) ($\Delta G = -9.4$ kcal/mol)

is only slightly uphill in enthalpy (by 3 kcal/mol) but it is in fact downhill in free enthalpy at room temperature conditions. It may be suspected that this fragility of $PrF_5$ in vicinity of other such molecules is the key reason beyond the failure to observe this molecule in experiments.

Related $PrF_4^+$ suffers from a similar fate. While unimolecular decomposition pathways are all uphill in both enthalpy and free enthalpy:

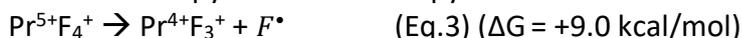
$Pr^{5+}F_4^+ \rightarrow Pr^{4+}F_3^+ + F^\bullet$       (Eq.3) ($\Delta G = +9.0$ kcal/mol)

the related bimolecular channel:

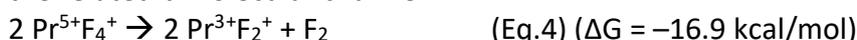
$2\,Pr^{5+}F_4^+ \rightarrow 2\,Pr^{3+}F_2^+ + F_2$       (Eq.4) ($\Delta G = -16.9$ kcal/mol)

is downhill in both factors. Its relative lack of stability as compared to $PrF_5$ is an exemplification of an old rule of chemistry which says that: "*it is easier to stabilize high OSs in anions, more difficult in neutral species, and most difficult in cations*".

Let us turn to O- and both O and F-containing species. For the $Pr^{5+}F_2O_2^-$ anion, which had singlet ground state at, all decomposition channels are uphill in both enthalpy and free enthalpy. The kinetics-limiting bimolecular channel corresponds to the process:

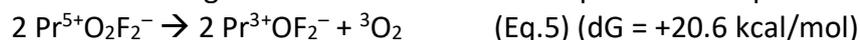
2 $Pr^{5+}O_2F_2^- \rightarrow$ 2 $Pr^{3+}OF_2^- + ^3O_2$        (Eq.5) (dG = +20.6 kcal/mol)

On the other hand, neutral $Pr^{5+}OF_3$ is uphill in dH0 and dG0 for all unimolecular processes, but one of bimolecular processes is strongly downhill:

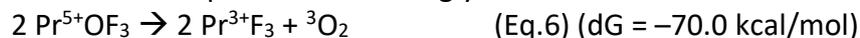
2 $Pr^{5+}OF_3 \rightarrow$ 2 $Pr^{3+}F_3 + ^3O_2$            (Eq.6) (dG = −70.0 kcal/mol)

Finally, their cationic relative, $Pr^{5+}OF_2^+$, one of the bimolecular processes is strongly downhill, causing the instability of that system.

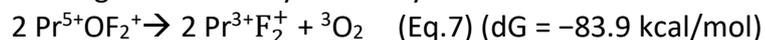
2 $Pr^{5+}OF_2^+ \rightarrow$ 2 $Pr^{3+}F_2^+ + ^3O_2$    (Eq.7) (dG = −83.9 kcal/mol)

$PrO_2^+$ is the only F-free cation in the molecular family studied here. The channel limiting is decomposition is uphill in free energy:

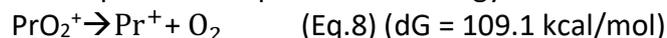
$PrO_2^+ \rightarrow Pr^+ + O_2$        (Eq.8) (dG = 109.1 kcal/mol)

However, the bimolecular channel is marginally downhill:

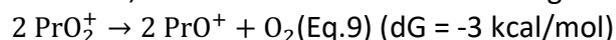
2 $PrO_2^+ \rightarrow$ 2 $PrO^+ + O_2$ (Eq.9) (dG = -3 kcal/mol)

PrNO is the only N-containing system for which calculations at the B3LYP/TZ2P level of theory predict a singlet ground state. All unimolecular channels are uphill, including:

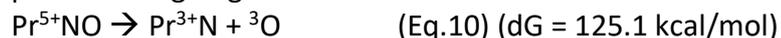
$Pr^{5+}NO \rightarrow Pr^{3+}N + ^3O$        (Eq.10) (dG = 125.1 kcal/mol)

and the bimolecular one:

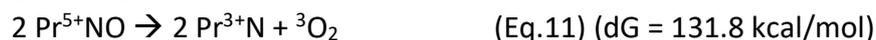
2 $Pr^{5+}NO \rightarrow$ 2 $Pr^{3+}N + ^3O_2$            (Eq.11) (dG = 131.8 kcal/mol)

is also uphill. This result correlates well with the successful observation of PrNO in experiment.

The bimolecular channels considered here do not account for a possible formation of dimers, e.g. reaction described by Eq.10 ends up in two $Pr^{3+}N$ molecular products rather than a dimer, $(Pr^{3+}N)_2$. Formation of dimers is expected to be particularly favored in the case of non-charged reaction products and will lead to further increase of exoergicity of decomposition reactions. The associated energy barriers for bimolecular processes are yet to be determined.

Concluding this section, we may state that in most cases the factor limiting thermal and thermodynamic stability of $Pr^{5+}$-bearing molecules is a possible bimolecular reaction. Only in one case, that of $Pr^{5+}F_2O_2^-$ anion, such process is predicted to be uphill in free enthalpy. This implies that high-OS derivatives of Pr should be sought after at very low temperature and large dilution conditions. In such conditions, bimolecular decay channels become irrelevant.

## 7. Property trends.

Trying to understand stability of species studied here it is worthwhile to arrange them in certain series, where clear property trends might emerge. As far as ligand substitution/removal and isoelectronic principle is concerned ($F^- \cong O^{2-} \cong N^{3-}$) one could arrange all species according to molecular charge and number of ligands (Table 9).

(a) The first generalization is that nearly all N-containing molecules, independent of their number of ligands and/or charge, do not correspond to genuine $Pr^{5+}$ systems. This means that the use of $N^{3-}$ ligand, which proved successful for TM compounds of high-valent $Ir^{9+}$ and $Pt^{10+}$, cannot be extended to Ln series which ends up at $Pr^{5+}$; here $Ce^{4+}$ is the likely limit. The only exception from this rule is PrNO, for which singlet is marginally stable with respect to triplet at B3LYP/TZ2P level, but the SO calculation tends to favor triplet-like geometry over singlet. We note that these authors (Ref.12) have obtained similarly contradictory results with DFT pointing slightly to triplet ground state while CCSD(T) ones to singlet one. Since this species was observed experimentally and confirmed to contain pentavalent lanthanide, the non-SO calculation in this case seems to reflect experimental reality better.

*Table 9. Summary of results for species studied in this work arranged with respect to the number of ligands and molecular charge. In **bold** font: molecules for which SO calculations predict a ground state corresponding to a genuine $Pr^{5+}$ species, as indicated by molecular geometry, BO analysis, and vibrational analysis. In Italics: molecules for which all considered bimolecular decomposition channels are uphill in standard free enthalpy, $\Delta G^0$, at SO level of theory. Underlined: molecules for which all considered bimolecular decomposition channels are uphill in standard enthalpy, $\Delta H^0$, at SO level of theory. For formulas in brackets cf. text as qualitative conclusions have been drawn in this section. Note, $PrN_2^-$ is computed to spontaneously detach electron in vacuum.*

| No ligands / charge | +1 | 0 | −1 |
|---|---|---|---|
| 2 | $PrNF^+$ $PrO_2^+$** | *$PrNO$*** | $PrN_2^-$ |
| 3 | **$PrOF_2^+$** | *$PrNF_2$* | [$PrNOF^-$] |
| 4 | **$PrF_4^+$** | **$PrOF_3$**[14] | *$PrNF_3^-$* *$\underline{PrO_2F_2^-}$* |
| 5 | --- | **$\underline{PrF_5}$***[10] | [$PrOF_4^-$] |
| 6 | --- | --- | [$PrF_6^-$] |

\* previously hypothesized \*\* previously prepared in experiment (Refs. 11, 12).

(b) The second observation is that among all O-containing species, the anionic one, $PrO_2F_2^-$, exhibits the most uphill decay channels; this agrees with the rule of thumb that high oxidation states are most easily preparable in anionic form. Thus, isoelectronic substitution $O^{2-} \rightarrow F^-$, which gradually decreases negative charge on molecular species in Series V, leads to decrease of stability.

Based on those, projections (extrapolations) could be made about [$PrF_6^-$], [$PrF_4O^-$] as well as [$PrNOF^-$] anions (none of which has been computed here).

1. Extension of generalization (a) to [$PrNOF^-$] anion suggests that also this molecule would not host genuine $Pr^{5+}$, i.e. its ground state would not be a singlet.
2. Extrapolation of observation (b) to [$PrOF_4^-$], together with the fact that analogous neutral species with five ligands ($PrF_5$) was successfully prepared, allows one to think that quest for $PrOF_4^-$ is worthwhile.
3. [$PrF_6^-$] is the only species from Table 9, which has as many as six ligands around $Pr^{5+}$; we note, that isostructural $V^{5+}$ most often exhibits either a tetrahedral or a distorted 5+1 coordination, with one ligand farther separated from metal; simply, there is not enough room around small pentavalent cations to host six (even small) ligands where Pauli repulsion of their lone pairs is crucial factor. Thus, stability of $PrF_6^-$ is unlikely. In some ways this is a common problem for all high valent species (e.g., $OsO_4$ is stable but $OsF_8$ is not).

**Conclusions**

We have studied using theoretical calculations a dozen of species which might contain pentavalent praseodymium. Aside from those already observed in experiment, our study reveals conceivable existence of $PrF_4^+$, $PrO_2F_2^-$ and $PrOF_2^+$. For each species we have calculated vibrational fundamentals for their facile detection in forthcoming experiments. Diverse stability limiting factors were discussed including uni- and bimolecular reactions which could lead to the decay of the most promising synthetic targets. The $Pr^{5+}$ species should be sought at very low temperature conditions and in high dilution to avoid decay via bimolecular channels. Extrapolation from quite stable (although somewhat oxidizing) $Ce^{4+}$ species via extremely fragile $Pr^{5+}$ ones, towards hypothetical $Nd^{6+}$ systems, suggests that the latter should not correspond to local minima.

**Acknowledgements**

The authors thank the Polish National Science Center for the OPUS project 2016/21/B/ST4/03866. ADF calculations have been carried out in Wroclaw Centre for Networking and Supercomputing (http://www.wcss.pl), using grant No. 367.7, as well as using Poland's high-performance Infrastructure PLGrid (HPC Center: WCSS) within computational grants no. PLG/2023/016309 and PLG/2022/015847. Gaussian calculations for several reference systems (Table 2) were performed at ICM, University of Warsaw, using grant No. G29-3.

**Keywords**

lanthanides ; praseodymium ; high oxidation state ; quantum mechanical calculations

**TOC Synopsis**

Based on theoretical calculations, $PrF_4^+$, $PrO_2F_2^-$ and $PrOF_2^+$ have been proposed as three new species which contain genuine pentavalent praseodymium, substantially enriching the small family of these exotic species.

**TOC graphics**

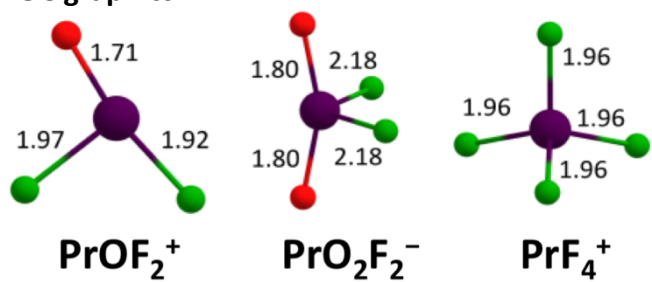

$PrOF_2^+$  $PrO_2F_2^-$  $PrF_4^+$

# Supplementary Materials

**Limits of stability for compounds of pentavalent praseodymium**

Piotr G. Szkudlarek[1], Paweł Szarek[1,2]*, Wojciech Grochala[1]*

[1] Center of New Technologies, University of Warsaw, Zwirki i Wigury 93, 02-089 Warsaw, Poland

[2] Navi-Chem, Włodarzewska 83/120, 02-393 Warsaw, Poland

S 1 Cartesian coordinates of all molecules optimized.

| Mult. | | Optimized geometry | | | Mult. | | Optimized geometry | | |
|---|---|---|---|---|---|---|---|---|---|
| | | $PrN_2^-$ | | | | | $PrOF_2^+$ | | |
| Spinorbit | Pr | -0.067418 | 0.685394 | 0.598723 | Spinorbit | Pr | -0.188302 | 0.284990 | 0.000128 |
| | N | 1.747390 | 0.828388 | 0.637968 | | O | 1.231178 | 1.234892 | 0.002215 |
| | N | -1.882315 | 0.544291 | 0.556772 | | F | 0.657042 | -1.489801 | -0.000147 |
| | | | | | | F | -1.967466 | -0.450103 | -0.002193 |
| Singlet | Pr | 0.000000 | 0.000000 | 0.000000 | Singlet | Pr | -0.180833 | 0.288986 | 0.002369 |
| | N | 1.718810 | 0.356458 | 0.000000 | | O | 1.249994 | 1.219870 | 0.001273 |
| | N | -1.718810 | -0.356458 | -0.000000 | | F | 0.630656 | -1.502165 | -0.009555 |
| | | | | | | F | -1.967368 | -0.426718 | 0.005913 |
| Triplet | Pr | -0.067447 | 0.686025 | 0.597821 | Triplet | Pr | -0.210880 | 0.219094 | 0.002143 |
| | N | 1.747117 | 0.828833 | 0.638794 | | O | 1.319329 | 1.291120 | 0.001340 |
| | N | -1.882012 | 0.543217 | 0.556848 | | F | 0.684527 | -1.521585 | -0.009845 |
| | | | | | | F | -2.060526 | -0.408655 | 0.006362 |
| Quintet | Pr | 0.013284 | 0.691997 | 0.599432 | Quintet | Pr | -0.425937 | -0.328012 | -0.775901 |
| | N | 1.780087 | 0.822699 | 0.634687 | | O | 1.378570 | 1.544197 | 0.257827 |
| | N | -1.995713 | 0.543379 | 0.559344 | | F | 0.812722 | -1.595190 | 0.147273 |
| | | | | | | F | -2.032908 | -0.041020 | 0.370801 |
| | | PrNO | | | | | $PrO_2F_2^-$ | | |
| Spinorbit | Pr | -0.018343 | -0.000000 | 0.000000 | Spinorbit | Pr | 0.042773 | 0.114442 | -0.000017 |
| | N | 1.762704 | -0.000000 | 0.000000 | | F | -0.950239 | 0.926794 | 1.747072 |
| | O | -1.832362 | 0.000000 | -0.000000 | | F | -0.950327 | 0.926191 | -1.747366 |
| | | | | | | O | 1.863546 | 0.380340 | -0.000037 |
| | | | | | | O | -0.289204 | -2.079477 | 0.000349 |
| Singlet | Pr | -0.058651 | -0.862241 | 0.318868 | Singlet | Pr | -0.075011 | -0.228738 | -0.000076 |
| | N | -1.740314 | -0.849716 | 0.284120 | | F | 2.043711 | 0.271185 | 0.001066 |
| | O | 1.715488 | -0.875454 | 0.355527 | | F | -1.487789 | 1.427438 | -0.000680 |
| | | | | | | O | -0.179311 | -0.544609 | 1.770601 |
| | | | | | | O | -0.177462 | -0.544313 | -1.770912 |
| Triplet | Pr | -0.039739 | -0.862131 | 0.319205 | Triplet | Pr | 0.017060 | 0.049493 | 0.113317 |
| | N | -1.803572 | -0.812021 | 0.274882 | | F | 2.079932 | 0.717440 | 0.021113 |
| | O | 1.759832 | -0.913256 | 0.364426 | | F | -1.258952 | 1.800755 | 0.023593 |
| | | | | | | O | -0.339239 | -1.037300 | 1.553836 |
| | | | | | | O | -0.374662 | -1.149424 | -1.711859 |
| Quintet | Pr | 0.191889 | -1.270705 | -0.227042 | Quintet | Pr | 0.035913 | 0.110531 | -0.006849 |
| | N | -1.958196 | -0.622066 | 0.524496 | | F | 2.095506 | 0.785616 | 0.007519 |
| | O | 1.682828 | -0.694635 | 0.661059 | | F | -1.230678 | 1.869373 | 0.007304 |
| | | | | | | O | -0.385203 | -1.184619 | 1.819333 |
| | | | | | | O | -0.391398 | -1.199936 | -1.827308 |
| | | PrNF⁺ | | | | | $PrF_3O$ | | |
| Spinorbit | Pr | -0.168623 | 0.841322 | 0.327213 | Spinorbit | Pr | 0.151800 | 0.097556 | 0.210865 |
| | F | 1.787571 | 1.103440 | 0.365584 | | F | 1.884088 | -0.644829 | -0.475715 |
| | N | -1.913395 | 0.608233 | 0.299372 | | F | -0.753991 | -1.730631 | 0.222859 |
| | | | | | | F | -0.202197 | 1.167308 | -1.489889 |
| | | | | | | O | -0.995611 | 0.928756 | 1.248555 |
| Singlet | Pr | 0.606719 | 0.069431 | -0.001790 | Singlet | Pr | -0.171546 | -0.057748 | 0.210183 |
| | N | 0.849785 | 1.698566 | 0.005353 | | F | 1.812352 | -0.543822 | 0.226770 |
| | F | 0.319705 | -1.854261 | -0.010224 | | F | -0.715244 | -1.846982 | -0.512581 |
| | | | | | | F | -0.766988 | 0.927882 | -1.476953 |
| | | | | | | O | -0.074489 | 1.338828 | 1.269256 |
| Triplet | Pr | 0.603799 | 0.042816 | -0.001386 | Triplet | Pr | -0.037194 | -0.006444 | -0.022920 |
| | N | 0.860139 | 1.771858 | 0.005418 | | F | 1.946954 | -0.408203 | -0.080990 |
| | F | 0.312271 | -1.900938 | -0.010693 | | F | -0.953346 | -1.828950 | -0.071520 |
| | | | | | | F | -0.426035 | 1.018926 | -1.743611 |
| | | | | | | O | -0.446294 | 1.042832 | 1.635717 |
| Quintet | Pr | 1.168741 | -0.170866 | -0.426937 | Quintet | Pr | 0.201828 | -0.436655 | -0.040324 |
| | N | 0.570097 | 1.757512 | 0.210017 | | F | 2.273236 | -0.624320 | -0.213859 |
| | F | 0.037372 | -1.672911 | 0.210257 | | F | -1.003846 | -2.136547 | -0.153787 |
| | | | | | | F | -0.618375 | 1.301783 | -1.025180 |
| | | | | | | O | -0.768756 | 1.713900 | 1.149828 |
| **PrNF$_2$** | | | | | **PrF$_4^+$** | | | | |
| Spinorbit | Pr | -0.051753 | 0.168777 | 0.000198 | Spinorbit | Pr | 0.000592 | 0.002153 | -0.000001 |
| | N | 1.175016 | 1.525649 | 0.002351 | | F | 1.873260 | 0.574115 | 0.000013 |
| | F | 0.663120 | -1.776566 | -0.000361 | | F | -1.168307 | 1.573064 | -0.000011 |
| | F | -2.053932 | -0.337882 | -0.002185 | | F | -0.352760 | -1.074667 | 1.596650 |
| | | | | | | F | -0.352785 | -1.074662 | -1.596651 |
| Singlet | Pr | -0.078486 | 0.235318 | 0.001702 | Singlet | Pr | -0.000261 | 0.001586 | -0.000596 |
| | N | 1.179140 | 1.366077 | 0.002261 | | F | 1.872619 | 0.574491 | 0.000301 |
| | F | 0.675563 | -1.687756 | -0.010625 | | F | -1.167685 | 1.574242 | 0.000770 |
| | F | -2.043768 | -0.333664 | 0.006661 | | F | -0.352390 | -1.074567 | 1.597408 |
| | | | | | | F | -0.352285 | -1.075751 | -1.597883 |
| Triplet | Pr | -0.032529 | 0.158499 | 0.000102 | Triplet | Pr | 0.282931 | -0.251115 | 0.185168 |
| | N | 1.179118 | 1.534634 | 0.001230 | | F | 2.135416 | 0.396225 | 0.053917 |
| | F | 0.631227 | -1.804686 | -0.000124 | | F | -1.131864 | 1.272018 | -0.529972 |
| | F | -2.045365 | -0.308475 | -0.001208 | | F | -0.331027 | -1.038590 | 1.879512 |
| | | | | | | F | -0.955456 | -0.378537 | -1.588625 |
| Quintet | Pr | -0.119095 | -0.187910 | 0.429669 | Quintet | Pr | -0.931497 | -0.119847 | 0.022886 |
| | N | 1.080373 | 1.703733 | -0.153535 | | F | 1.337477 | -1.228341 | -0.005203 |
| | F | 0.879599 | -1.925610 | -0.127399 | | F | -1.966190 | 1.570643 | 0.012445 |
| | F | -2.108425 | -0.010238 | -0.148737 | | F | 0.813093 | -0.081562 | 1.479948 |
| | | | | | | F | 0.747115 | -0.140896 | -1.510078 |
| **PrNF$_3^-$** | | | | | **PrF$_5$** | | | | |
| Spinorbit | Pr | -0.192750 | -0.009877 | -0.000008 | Spinorbit | Pr | 0.000011 | -0.000010 | -0.000057 |
| | N | 2.283800 | 0.102741 | 0.000160 | | F | 1.690238 | 0.059386 | 1.053986 |
| | F | -0.747691 | -2.099285 | -0.000086 | | F | 0.626619 | -1.689465 | -0.909752 |
| | F | -0.920217 | 0.990032 | -1.783442 | | F | 0.437053 | 1.799980 | -0.802232 |
| | F | -0.920442 | 0.989963 | 1.783375 | | F | -1.690327 | -0.099656 | -1.053912 |
| | | | | | | F | -1.063594 | -0.110235 | 1.711970 |
| Singlet | Pr | 0.121677 | 0.006449 | 0.001841 | Singlet | Pr | 0.380059 | -0.000058 | 0.000012 |
| | N | 1.857851 | 0.098808 | 0.000840 | | F | 2.208745 | 0.555031 | 0.651599 |
| | F | -0.723009 | -1.976477 | 0.000798 | | F | -0.998153 | 1.404086 | -0.453601 |
| | F | -0.878073 | 0.923752 | -1.672835 | | F | -0.070040 | -1.958388 | -0.197998 |
| | F | -0.875746 | 0.921041 | 1.679357 | | F | -0.278488 | -0.039146 | 1.881169 |
| | | | | | | F | 1.038540 | 0.038470 | -1.881184 |
| Triplet | Pr | 0.056928 | 0.003186 | 0.001929 | Triplet | Pr | 0.129961 | -0.059191 | 0.098695 |
| | N | 1.975365 | 0.107672 | 0.000734 | | F | 2.329367 | 0.515879 | 0.243090 |
| | F | -0.737452 | -1.995126 | 0.000625 | | F | -1.277393 | 1.242036 | -0.527855 |
| | F | -0.897212 | 0.930156 | -1.687658 | | F | -0.443895 | -1.942210 | -0.338365 |
| | F | -0.894928 | 0.927684 | 1.694369 | | F | -0.099854 | 0.002052 | 2.101084 |
| | | | | | | F | 1.642487 | 0.241434 | -1.576649 |
| Quintet | Pr | -0.208873 | -0.010776 | -0.000052 | Quintet | Pr | 0.174535 | 0.237451 | -0.045226 |
| | N | 2.231784 | 0.112415 | 0.001116 | | F | 2.404583 | 0.820347 | -0.161474 |
| | F | -0.735484 | -2.112784 | -0.000284 | | F | -1.164611 | 1.783499 | -0.338730 |
| | F | -0.891504 | 0.992345 | -1.795450 | | F | -0.768771 | -1.830525 | 0.355055 |
| | F | -0.893221 | 0.992375 | 1.794669 | | F | -0.045394 | -0.901165 | 1.950154 |
| | | | | | | F | 1.680331 | -0.109605 | -1.759778 |
| **PrO$_2^+$** | | | | | | | | | |
| Spinorbit | Pr | -0.000000 | -0.000000 | -0.000000 | | | | | |
| | O | 1.722753 | 0.357288 | 0.000008 | | | | | |
| | O | -1.722752 | -0.357288 | -0.000008 | | | | | |

| Singlet | Pr | -0.000000 | -0.000000 | -0.000000 |
|---------|----|-----------|-----------|-----------|
|         | O  | 1.660963  | 0.344464  | 0.000003  |
|         | O  | -1.660963 | -0.344464 | -0.000003 |
| Triplet | Pr | 0.000000  | 0.000000  | 0.000000  |
|         | O  | 1.709920  | 0.354614  | 0.000000  |
|         | O  | -1.709920 | -0.354614 | -0.000000 |
| Quintet | Pr | -0.150550 | 0.810200  | -0.001889 |
|         | O  | 1.702396  | -0.104449 | 0.001012  |
|         | O  | -1.551845 | -0.705750 | 0.000877  |

S 2 Frequencies for all analyzed systems in all possible multiplicities.

| Multiplicity | Frequency [cm$^{-1}$] | Assign. | Multiplicity | Frequency [cm$^{-1}$] | Assign. |
|---|---|---|---|---|---|
| **PrN$_2^-$** | | | **PrNF$_2$** | | |
| Spinorbit | 167 | $\delta_{N-Pr-N}$ | Spinorbit | 499 | $\nu_{as\,F-Pr-F}$ |
| | 742 | $\nu_{s\,N-Pr-N}$ | | 543 | $\nu_{s\,F-Pr-F}$ |
| | 749 | $\nu_{as\,N-Pr-N}$ | | 718 | $\nu_{Pr-N}$ |
| Singlet | 129, 132 | $\delta_{N-Pr-N}$ | Singlet | 116 | $\delta_{N-Pr-F}$ |
| | 860 | $\nu_{s\,N-Pr-N}$ | | 119 | $\pi_{N-Pr-F}$ |
| | 907 | $\nu_{as\,N-Pr-N}$ | | 133 | $\delta_{F-Pr-F}$ |
| Triplet | 166,167 | $\delta_{N-Pr-N}$ | | 505 | $\nu_{as\,F-Pr-F}$ |
| | 743 | $\nu_{s\,N-Pr-N}$ | | 557 | $\nu_{s\,F-Pr-F}$ |
| | 748 | $\nu_{as\,N-Pr-N}$ | | 954 | $\nu_{Pr-N}$ |
| Quintet | 74, 133 | $\delta_{N-Pr-N}$ | Triplet | 49 | $\rho_{Pr-F}$ |
| | 433 | $\nu_{as\,N-Pr-N}$ | | 94 | $\pi_{N-Pr-F}$ |
| | 799 | $\nu_{s\,N-Pr-N}$ | | 130 | $\delta_{F-Pr-F}$ |
| **PrNO** | | | | 499 | $\nu_{as\,F-Pr-F}$ |
| Spinorbit | 163 | $\delta_{N-Pr-O}$ | | 544 | $\nu_{s\,F-Pr-F}$ |
| | 737 | $\nu_{Pr-O}$ | | 724 | $\nu_{Pr-N}$ |
| | 815 | $\nu_{Pr-N}$ | Quintet | 71, 76 | $\delta_{N-Pr-F}$ |
| Singlet | 158 | $\delta_{O-Pr-N}$ | | 110 | $\delta_{F-Pr-F}$ |
| | 822 | $\nu_{Pr-O}$ | | 336 | $\nu_{Pr-N}$ |
| | 1029 | $\nu_{Pr-N}$ | | 516 | $\nu_{as\,Pr-F}$ |
| Triplet | 140, 152 | $\delta_{N-Pr-O}$ | | 532 | $\nu_{Pr-F}$ |
| | 748 | $\nu_{Pr-O}$ | **PrNF$_3^-$** | | |
| | 825 | $\nu_{Pr-N}$ | Spinorbit | 105 | $\tau$ |
| Quintet | 87 | $\delta_{N-Pr-O}$ | | 111 | $\pi_{F-Pr-F}$ |
| | 350 | $\nu_{Pr-N}$ | | 114 | $\delta_{N-Pr-F}$ |
| | 805 | $\nu_{Pr-O}$ | | 119 | $\delta_{F-Pr-F\,+\,N-Pr-F}$ |
| **PrNF$^+$** | | | | 364 | $\nu_{Pr-N}$ |
| Spinorbit | 115 | $\delta_{N-Pr-F}$ | | 442, 443 | $\nu_{as\,F-Pr-F}$ |
| | 604 | $\nu_{Pr-F}$ | | 472 | $\nu_{s\,Pr-F\,+\,Pr-N}$ |
| | 849 | $\nu_{Pr-N}$ | Singlet | 100 | $\tau$ |
| Singlet | 131, 133 | $\delta_{F-Pr-N}$ | | 100 | $\delta_{F-Pr-N}$ |
| | 619 | $\nu_{Pr-F}$ | | 121 | $\delta_{N-Pr-F}$ |
| | 1068 | $\nu_{Pr-N}$ | | 132 | $\pi_{F-Pr-F}$ |
| Triplet | 155 | $\delta_{F-Pr-N}$ | | 437 | $\nu_{as\,F-Pr-F}$ |
| | 603 | $\nu_{Pr-F}$ | | 480 | $\nu_{s\,F-Pr-F}$ |
| | 864 | $\nu_{Pr-N}$ | | 859 | $\nu_{Pr-N}$ |
| Quintet | 142 | $\delta_{F-Pr-N}$ | Triplet | 97 | $\tau$ |
| | 582 | $\nu_{Pr-F}$ | | 126 | $\delta_{N-Pr-F}$ |
| | 624 | $\nu_{Pr-N}$ | | 132 | $\pi_{N-Pr-F}$ |
| **PrNF$_2$** | | | | 441 | $\nu_{as\,F-Pr-F}$ |
| Spinorbit | 60 | $\rho_{N-Pr-F}$ | | 471 | $\nu_{s\,Pr-F}$ |
| | 99 | $\pi_{N-Pr-F}$ | | 634 | $\nu_{Pr-N}$ |
| | 130 | $\delta_{F-Pr-F}$ | **Quintet** | | |

| Multiplicity | Frequency [cm⁻¹] | Assign. | Multiplicity | Frequency [cm⁻¹] | Assign. |
|---|---|---|---|---|---|
| $PrNF_3^-$ | | | $PrNF_3^-$ | | |
| Quintet | 108 | $\tau$ | | 591 | $\nu_{as\ F-Pr-F}$ |
| | 115 | $\delta_{N-Pr-F}$ | | 625 | $\nu_{s\ F-Pr-F}$ |
| | 117, 118 | $\pi_{N-Pr-F}$ | $PrO_2F_2^-$ | | |
| | 375 | $\nu_{Pr-N}$ | Spinorbit | 103 | $\tau$ |
| | 440 | $\nu_{as\ F-Pr-F}$ | | 111 | $\delta_{F-Pr-F}$ |
| | 469 | $\nu_{s\ Pr-N\ \&\ Pr-F}$ | | 120 | $\delta_{O-Pr-F+O-Pr-O}$ |
| $PrO_2^+$ | | | | 143 | $\delta_{O-Pr-O}$ |
| Spinorbit | 195, 199 | $\delta_{O-Pr-O}$ | | 144 | $\pi_{O-Pr-O}$ |
| | 755 | $\nu_{s\ O-Pr-O}$ | | 393 | $\nu_{Pr-O}$ |
| | 783 | $\nu_{as\ O-Pr-O}$ | | 442 | $\nu_{as\ F-Pr-F}$ |
| Singlet | 202,206 | $\delta_{O-Pr-O}$ | | 459 | $\nu_{s\ F-Pr-F}$ |
| | 884 | $\nu_{s\ O-Pr-O}$ | | 759 | $\nu_{Pr-O}$ |
| | 1017 | $\nu_{as\ O-Pr-O}$ | Singlet | 82 | $\delta_{F-Pr-F}$ |
| Triplet | 200, 202 | $\delta_{O-Pr-O}$ | | 178 | $\pi_{F-Pr-F}$ |
| | 754 | $\nu_{s\ O-Pr-O}$ | | 193 | $\delta_{O-Pr-O}$ |
| | 805 | $\nu_{as\ O-Pr-O}$ | | 196 | $\tau$ |
| Quintet | 108 | $\delta_{O-Pr-O}$ | | 237 | $\pi_{O-Pr-O}$ |
| | 510 | $\nu_{as\ O-Pr-O}$ | | 430 | $\nu_{s\ F-Pr-F}$ |
| | 560 | $\nu_{s\ O-Pr-O}$ | | 432 | $\nu_{as\ F-Pr-F}$ |
| $PrOF_2^+$ | | | | 731 | $\nu_{s\ O-Pr-O}$ |
| Spinorbit | 156 | $\pi_{O-Pr-F}$ | | 806 | $\nu_{as\ O-Pr-O}$ |
| | 160 | $\delta_{F-Pr-F}$ | Triplet | 80 | $\delta_{O-Pr-F}$ |
| | 181 | $\delta_{F-Pr-O}$ | | 106 | $\tau$ |
| | 599 | $\nu_{as\ F-Pr-F}$ | | 113 | $\delta_{F-Pr-F}$ |
| | 649 | $\nu_{s\ F-Pr-F}$ | | 120 | $\pi_{F-Pr-F}$ |
| | 900 | $\nu_{Pr-O}$ | | 143 | $\delta_{O-Pr-O}$ |
| Singlet | 156 | $\pi_{O-Pr-F}$ | | 396 | $\nu_{as\ F-Pr-F}$ |
| | 161 | $\delta_{F-Pr-F}$ | | 402 | $\nu_{Pr-O}$ |
| | 180 | $\delta_{F-Pr-O}$ | | 458 | $\nu_{s\ F-Pr-F}$ |
| | 598 | $\nu_{s\ F-Pr-F}$ | | 759 | $\nu_{Pr-O}$ |
| | 649 | $\nu_{s\ F-Pr-F}$ | Quintet | 100 | $\tau$ |
| | 901 | $\nu_{as\ F-Pr-O}$ | | 113 | $\delta_{O-Pr-O}$ |
| Triplet | 116 | $\pi_{O-Pr-F}$ | | 116 | $\pi_{O-Pr-O}$ |
| | 136 | $\delta_{F-Pr-F}$ | | 120 | $\pi_{F-Pr-O}$ |
| | 147 | $\delta_{O-Pr-F}$ | | 124 | $\delta_{F-Pr-F}$ |
| | 538 | $\nu_{as\ F-Pr-F}+\nu_O$ | | 410 | $\nu_{as\ O-Pr-O}$ |
| | 631 | $\nu_{O-Pr-F}$ | | 423 | $\nu_{s\ O-Pr-O\ \&\ F-Pr-F}$ |
| | 641 | $\nu_{s\ F-Pr-F}$ | | 433 | $\nu_{as\ F-Pr-F}$ |
| Quintet | 29 | $\rho_{F-Pr-F}$ | | 467 | $\nu_{Pr-F\ \&\ Pr-O}$ |
| | 35 | $\pi_{F-Pr-F}$ | $PrF_3O$ | | |
| | 123 | $\delta_{F-Pr-F}$ | Spinorbit | 77 | $\delta_{F-Pr-F}$ |
| | 183 | $\nu_{Pr-O}$ | | 143 | $\delta_{O-Pr-F}$ |

| Multiplicity | Frequency [cm⁻¹] | Assign. | Multiplicity | Frequency [cm⁻¹] | Assign. |
|---|---|---|---|---|---|
| | **PrF₃O** | | | **PrF₄⁺** | |
| | 158 | $\delta_{F-Pr-F}$ | Singlet | 622 | $\nu_{s\,F-Pr-F}$ |
| | 166 | $\pi_{F-Pr-F}$ | Triplet | 29 | $\delta_{F-Pr-F}$ |
| | 193 | $\delta_{O-Pr-F}$ | | 67 | $\pi_{F-Pr-F}$ |
| | 518 | $\nu_{as\,F-Pr-F}$ | | 71 | $\tau$ |
| | 548 | $\nu_{as\,F-Pr-F}$ | | 97 | $\pi_{F-Pr-F}$ |
| | 584 | $\nu_{s\,F-Pr-F}$ | | 196 | $\nu_{as\,F-Pr-F}$ |
| | 815 | $\nu_{Pr-O}$ | | 343 | $\delta_{F-Pr-F}$ |
| Singlet | 77 | $\delta_{F-Pr-F}$ | | 510 | $\nu_{s\,F-Pr-F}$ |
| | 143 | $\delta_{O-Pr-F}$ | | 630 | $\nu_{s\,F-Pr-F}$ |
| | 159 | $\rho_{F-Pr-F}$ | | 664 | $\nu_{as\,F-Pr-F}$ |
| | 166 | $\pi_{F-Pr-F}$ | Quintet | 61 | $\pi_{F-Pr-F}$ |
| | 192 | $\delta_{O-Pr-F}$ | | 69 | $\delta_{F-Pr-F}$ |
| | 517 | $\nu_{as\,F-Pr-F}$ | | 83 | $\pi_{F-Pr-F}$ |
| | 546 | $\nu_{as\,F-Pr-F}$ | | 150 | $\delta_{F-Pr-F}$ |
| | 584 | $\nu_{s\,F-Pr-F}$ | | 255 | $\nu_{as\,F-Pr-F}$ |
| | 816 | $\nu_{Pr-O}$ | | 367 | $\nu_{as\,F-Pr-F}$ |
| Triplet | 97 | $\pi_{F-Pr-F}$ | | 421 | $\delta_{F-Pr-F}$ |
| | 110 | $\delta_{F-Pr-F}$ | | 447 | $\pi_{F-Pr-F}$ |
| | 121 | $\tau$ | | 624 | $\nu_{s\,Pr-F}$ |
| | 128 | $\delta_{O-Pr-F}$ | | **PrF₅** | |
| | 193 | $\delta_{O-Pr-F}$ | Spinorbit | 53 | $\delta_{F-Pr-F}$ |
| | 536 | $\nu_{Pr-O\,\&\,Pr-F}$ | | 156 | $\pi_{F-Pr-F}$ |
| | 556 | $\nu_{as\,F-Pr-F}$ | | 165 | $\delta_{F-Pr-F}$ |
| | 586 | $\nu_{s\,F-Pr-F\,\&\,F-Pr-O}$ | | 180 | $\tau$ |
| | 601 | $\nu_{as\,F-Pr-F}$ | | 492 | $\nu_{as\,F-Pr-F}$ |
| Quintet | 59 | $\tau$ | | 560 | $\nu_{as\,F-Pr-F}$ |
| | 62 | $\pi_{F-Pr-F}$ | | 585 | $\nu_{s\,F-Pr-F}$ |
| | 112 | $\delta_{F-Pr-F}$ | | 587 | $\nu_{as\,F-Pr-F}$ |
| | 124 | $\delta_{F-Pr-F}$ | Singlet | 54 | $\delta_{F-Pr-F}$ |
| | 200 | $\nu_{Pr-O\,\&\,Pr-F}$ | | 156 | $\pi_{F-Pr-F}$ |
| | 211 | $\delta_{O-Pr-F}$ | | 165 | $\delta_{F-Pr-F}$ |
| | 460 | $\nu_{Pr-F}$ | | 179, 180 | $\tau$ |
| | 518 | $\nu_{as\,F-Pr-F}$ | | 492 | $\nu_{as\,F-Pr-F}$ |
| | 533 | $\nu_{s\,F-Pr-F}$ | | 560 | $\nu_{as\,F-Pr-F}$ |
| | **PrF₄⁺** | | | 586 | $\nu_{s\,F-Pr-F}$ |
| Spinorbit | 43, 44, 45 | $\delta_{F-Pr-F}$ | | 586 | $\nu_{as\,F-Pr-F}$ |
| | 64 | $\tau$ | Triplet | 13 | $\tau$ |
| | 611 | $\nu_{as\,F-Pr-F}$ | | 88 | $\rho_{Pr-F}$ |
| | 622 | $\nu_{s\,F-Pr-F}$ | | 92 | $\pi_{F-Pr-F}$ |
| Singlet | 43, 44, 44 | $\delta_{F-Pr-F}$ | | 119 | $\delta_{F-Pr-F}$ |
| | 63, 64 | $\tau$ | | 132 | $\pi_{F-Pr-F}$ |
| | 611 | $\nu_{as\,F-Pr-F}$ | | 139 | $\delta_{F-Pr-F}$ |

| Multiplicity | Frequency [cm$^{-1}$] | Assign. |
|---|---|---|
| PrF$_5$ | | |
| Triplet | 237 | $\nu$ $_{as\ F\text{-}Pr\text{-}F}$ |
| | 335 | $\delta$ $_{F\text{-}Pr\text{-}F}$ |
| | 465 | $\nu$ $_{s\ F\text{-}Pr\text{-}F}$ |
| | 556 | $\nu$ $_{as\ F\text{-}Pr\text{-}F}$ |
| | 566 | $\nu$ $_{as\ F\text{-}Pr\text{-}F}$ |
| | 595 | $\nu$ $_{s\ F\text{-}Pr\text{-}F}$ |
| Quintet | 46 | $\rho$ $_{F\text{-}Pr\text{-}F}$ |
| | 56 | $\tau$ |
| | 76, 77 | $\pi$ $_{F\text{-}Pr\text{-}F}$ |
| | 81 | $\tau$ |
| | 254 | $\nu$ $_{as\ F\text{-}Pr\text{-}F}$ |
| | 259 | $\nu$ $_{as\ F\text{-}Pr\text{-}F}$ |
| | 313 | $\delta$ $_{F\text{-}Pr\text{-}F}$ |
| | 324 | $\nu$ $_{as\ F\text{-}Pr\text{-}F}$ |
| | 437 | $\nu$ $_{as\ F\text{-}Pr\text{-}F}$ |
| | 442 | $\delta$ $_{F\text{-}Pr\text{-}F}$ |
| | 549 | $\nu$ $_{s\ Pr\text{-}F}$ |

**S 3 Bond lengths and bond orders for all analyzed systems**

| Mult. | Bond | Distance [Å] | G-J | N-M | Mult. | Bond | Distance [Å] | G-J | N-M |
|---|---|---|---|---|---|---|---|---|---|
| $PrN_2^-$ | | | | | $PrOF_2^+$ | | | | |
| Singlet | Pr-N | 1.76 | 2.12 | 3.2 | Quintet | Pr-O | 2.80 | 0.11 | 0.82 |
| Triplet | Pr-N | 1.82 | 1.89 | 2.9 | | Pr-F | 2.00 | 0.61 | 1.94 |
| Quintet | Pr-N | 1.77 | 1.89 | 2.96 | | Pr-F | 1.99 | 0.62 | 1.96 |
| $PrNO$ | | | | | $PrF_2O_2^-$ | | | | |
| Singlet | Pr-N | 1.68 | 2.34 | 3.64 | Singlet | Pr-F | 2.18 | 0.41 | 0.92 |
| | Pr-O | 1.77 | 1.48 | 2.44 | | Pr-O | 1.80 | 1.52 | 2.42 |
| Triplet | Pr-N | 1.77 | 2.01 | 3.05 | Triplet | Pr-F | 2.17 | 0.39 | 0.97 |
| | Pr-O | 1.8 | 1.38 | 2.24 | | Pr-O | 1.84 | 1.54 | 2.85 |
| Quintet | Pr-N | 2.37 | 0.57 | 1.39 | Quintet | Pr-F | 2.17 | 0.35 | 1.22 |
| | Pr-O | 1.83 | 1.46 | 3.39 | | Pr-O | 2.28 | 0.4 | 1.58 |
| $PrNF^+$ | | | | | $PrF_3O$ | | | | |
| Singlet | Pr-N | 1.65 | 2.58 | 4.28 | | | | | |
| | Pr-F | 1.95 | 0.7 | 1.29 | Singlet | Pr-F | 2.04 | 0.65 | 1.18 |
| Triplet | Pr-N | 1.75 | 2.17 | 3.75 | | Pr-F | 2.00 | 0.67 | 1.14 |
| | Pr-F | 1.97 | 0.66 | 1.28 | | Pr-O | 1.76 | 1.85 | 2.8 |
| Quintet | Pr-N | 2.12 | 0.9 | 2.35 | Triplet | Pr-F | 2.03 | 0.63 | 1.3 |
| | Pr-F | 1.99 | 0.64 | 1.85 | | Pr-F | 2.04 | 0.6 | 1.25 |
| $PrNF_2$ | | | | | | Pr-O | 2.00 | 0.92 | 1.91 |
| Singlet | Pr-N | 1.69 | 2.52 | 3.91 | Quintet | Pr-F | 2.09 | 0.45 | 1.53 |
| | Pr-F | 2.07 | 0.58 | 1.15 | | Pr-F | 2.16 | 0.34 | 1.04 |
| | Pr-F | 2.05 | 0.56 | 1.06 | | Pr-O | 2.64 | 0.12 | 0.46 |
| Triplet | Pr-N | 1.83 | 2.02 | 3.33 | | F-O | 2.22 | 0.28 | 0.76 |
| | Pr-F | 2.07 | 0.52 | 1.11 | $PrF_4^+$ | | | | |
| Quintet | Pr-N | 2.31 | 0.6 | 1.69 | Singlet | Pr-F | 1.96 | 0.98 | 1.47 |
| | Pr-F | 2.08 | 0.46 | 1.46 | Triplet | Pr-F | 1.97 | 0.83 | 1.68 |
| $PrNF_3^-$ | | | | | | Pr-F | 2.2 | 0.36 | 0.73 |
| Singlet | Pr-N | 1.74 | 2.58 | 3.79 | | Pr-F | 2.17 | 0.42 | 0.83 |
| | Pr-F | 2.16 | 0.44 | 0.94 | | F-F | 1.97 | 0.46 | 0.71 |
| Triplet | Pr-N | 1.92 | 1.82 | 3.05 | Quintet | Pr-F | 2.53 | 0.1 | 0.3 |
| | Pr-F | 2.15 | 0.4 | 0.96 | | Pr-F | 1.98 | 0.64 | 2.15 |
| Quintet | Pr-N | 2.44 | 0.43 | 1.35 | | Pr-F | 2.27 | 0.23 | 0.74 |
| Quintet | Pr-F | 2.17 | 0.35 | 1.23 | | F-F | 1.95 | 0.48 | 0.73 |
| $PrO_2^+$ | | | | | | F-F | 1.95 | 0.48 | 0.73 |
| Singlet | Pr-O | 1.7 | 1.75 | 2.88 | $PrF_5$ | | | | |
| Triplet | Pr-O | 1.75 | 1.55 | 2.48 | | | | | |
| Quintet | Pr-O | 2.07 | 0.75 | 2.3 | Singlet | Pr-F | 2.02 | 0.78 | 1.26 |
| | Pr-O | 2.06 | 0.75 | 2.3 | | Pr-F | 1.99 | 0.82 | 1.28 |
| $PrOF_2^+$ | | | | | Triplet | Pr-F | 2.28 | 0.25 | 0.51 |
| Singlet | Pr-O | 1.71 | 1.94 | 3.02 | | Pr-F | 2.02 | 0.66 | 1.4 |
| | Pr-F | 1.97 | 0.86 | 1.48 | | Pr-F | 2.02 | 0.68 | 1.42 |
| | Pr-F | 1.92 | 0.81 | 1.35 | | Pr-F | 2.28 | 0.26 | 0.53 |
| Triplet | Pr-O | 1.87 | 1.31 | 2.36 | | F-F | 1.96 | 0.48 | 0.82 |
| | Pr-F | 1.96 | 0.79 | 1.5 | Quintet | Pr-F | 2.31 | 0.19 | 0.61 |
| | Pr-F | 1.95 | 0.75 | 1.38 | | Pr-F | 2.07 | 0.48 | 1.67 |
| | | | | | | F-F | 1.99 | 0.49 | 0.89 |
| | | | | | | F-F | 1.98 | 0.49 | 0.89 |

S 4 Mulliken and Hirschfeld populational analyses.

| Mult. | Mulliken | | | | Hirshfeld | | | |
|---|---|---|---|---|---|---|---|---|
| | Pr | N | O | F | Pr | N | O | F |
| PrN$_2^-$ | | | | | | | | |
| Spinorbit | 0.45 | −0.73 | - | - | 0.48 | −0.74 | - | - |
| Singlet | 0.34 | −0.67 | - | - | 0.50 | −0.75 | - | - |
| Triplet | 0.45 | −0.73 | - | - | 0.48 | −0.74 | - | - |
| Quintet | 0.20 | −0.60 | - | - | 0.15 | −0.63, −0.52 | - | - |
| PrNO | | | | | | | | |
| Spinorbit | 1.23 | −0.54 | −0.69 | - | 1.05 | −0.53 | −0.52 | - |
| Singlet | 1.16 | −0.49 | −0.66 | - | 1.08 | −0.58 | −0.50 | - |
| Triplet | 1.23 | −0.54 | −0.69 | - | 1.06 | −0.54 | −0.51 | - |
| Quintet | 1.22 | −0.52 | −0.70 | - | 0.93 | −0.38 | −0.56 | - |
| PrNF$^+$ | | | | | | | | |
| Spinorbit | 1.75 | −0.26 | - | −0.48 | 1.53 | −0.28 | - | −0.24 |
| Singlet | 1.72 | -0.26 | - | −0.47 | 1.57 | −0.34 | - | −0.22 |
| Triplet | 1.75 | −0.27 | - | −0.48 | 1.53 | −0.29 | - | −0.24 |
| Quintet | 1.78 | −0.27 | - | −0.51 | 1.45 | −0.17 | - | −0.28 |
| PrNF$_2$ | | | | | | | | |
| Spinorbit | 1.66 | −0.44 | - | −0.61, −0.61 | 1.14 | −0.41 | - | −0.37, −0.36 |
| Singlet | 1.6 | −0.41 | - | −0.60, −0.59 | 1.15 | −0.46 | - | −0.35, −0.34 |
| Triplet | 1.66 | −0.44 | - | −0.61, −0.61 | 1.14 | −0.41 | - | −0.37, −0.36 |
| Quintet | 1.73 | −0.46 | - | −0.63, −0.63 | 1.1 | −0.32 | - | −0.39 [x2] |
| PrNF$_3^-$ | | | | | | | | |
| Spinorbit | 1.76 | −0.62 | - | −0.71 [x3] | 0.83 | −0.44 | - | −0.46, −0.47 [x2] |
| Singlet | 1.58 | −0.53 | - | −0.69 [x3] | 0.85 | −0.54 | - | −0.44 [x3] |
| Triplet | 1.68 | −0.59 | - | −0.69 [x3] | 0.83 | −0.51 | - | −0.44 [x3] |
| Quintet | 1.75 | −0.61 | - | −0.71 [x3] | 0.81 | −0.43 | - | −0.46 [x3] |
| PrO$_2^+$ | | | | | | | | |
| Spinorbit | 1.89 | - | −0.44 | - | 1.55 | - | −0.28 | - |
| Singlet | 1.88 | - | −0.44 | - | 1.59 | - | −0.29 | - |
| Triplet | 1.90 | - | −0.45 | - | 1.56 | - | −0.28 | - |
| Quintet | 1.75 | - | −0.37 | - | 1.44 | - | −0.22 | - |
| PrOF$_2^+$ | | | | | | | | |
| Spinorbit | 2.21 | - | −0.36 | −0.44, −0.41 | 1.57 | - | −0.2 | −0.2, −0.16 |
| Singlet | 2.21 | - | −0.36 | −0.44, −0.41 | 1.57 | - | −0.2 | −0.2, −0.16 |
| Triplet | 2.21 | - | −0.32 | −0.45, −0.44 | 1.55 | - | −0.15 | −0.21, −0.19 |
| Quintet | 2.00 | - | 0.07 | −0.53 | 1.4 | - | 0.19 | −0.29 |
| PrF$_2$O$_2^-$ | | | | | | | | |
| Spinorbit | 1.74 | - | −0.72, −0.61 | −0.7 | 0.86 | - | −0.54, −0.41 | −0.45 |
| Singlet | 1.77 | - | −0.68 | −0.7 | 0.88 | - | −0.48 | −0.46 |
| Triplet | 1.74 | - | −0.72, −0.61 | −0.7 | 0.86 | - | −0.54, −0.41 | −0.45 |
| Quintet | 1.71 | - | −0.65, −0.64 | −0.71 | 0.8 | - | −0.44, −0.43 | −0.46 |
| PrF$_3$O | | | | | | | | |
| Spinorbit | 2.15 | - | −0.49 | −0.53, −0.56 | 1.21 | - | −0.31 | −0.27, −0.31[x2] |
| Singlet | 2.14 | - | −0.49 | −0.56, −0.53 | 1.21 | - | −0.31 | −0.31 [x2], −0.27 |
| Triplet | 2.15 | - | −0.45 | −0.56, −0.57 | 1.21 | - | −0.26 | −0.31, −0.32 [x2] |
| Quintet | 1.94 | - | −0.08 | −0.64, −0.57 | 1.1 | - | 0.03 | −0.40 [x2], −0.34 |
| PrF$_4^+$ | | | | | | | | |
| Spinorbit | 2.5 | - | - | −0.38 [x4] | 1.55 | - | - | −0.14 [x4] |
| Singlet | 2.5 | - | - | −0.37 [x4] | 1.55 | - | - | −0.14 [x4] |
| Triplet | 2.4 | - | - | −0.45 [x2], −0.24, −0.27 | 1.55 | | | −0.21 [x2], −0.05, −0.08 |
| Quintet | 2.15 | - | - | −0.09, −0.52, −0.27 [x2] | 1.43 | - | - | 0.02, −0.28, −0.09 [x2] |
| PrF$_5$ | | | | | | | | |
| Spinorbit | 2.54 | - | - | −0.52 [x2], −0.5 [x3] | 1.21 | - | - | −0.22 [x2], −0.26 [x3] |
| Singlet | 2.54 | - | - | −0.5 [x3], −0.52 [x2] | 1.21 | - | - | −0.26 [x3], −0.22 [x2] |
| Triplet | 2.34 | - | - | −0.34 [x2], −0.55 [x2], −0.56 | 1.2 | - | - | −0.15 [x2], −0.3 [x3] |
| Quintet | 2.08 | - | - | −0.36 [x4], −0.62 | 1.09 | - | - | −0.18 [x4], −0.37 |

| $PrN_2^-$ | | |
|---|---|---|
| Orbital | Isosurface | % Pr $f$ |
| HOMO-2 | 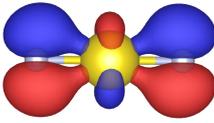 | 23.39 |
| HOMO-1 | 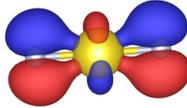 | 23.34 |
| HOMO | 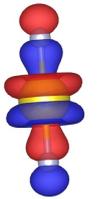 | 70.72 |
| LUMO+1 | 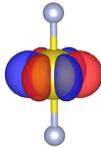 | 99.26 |
| LUMO+2 | 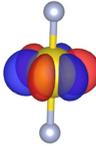 | 99.26 |
| LUMO+6 | 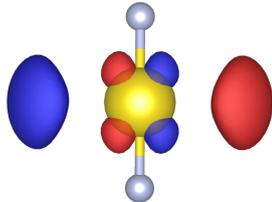 | 19.81 |
| LUMO+7 | 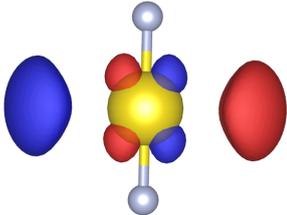 | 19.76 |
| LUMO+8 | 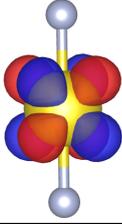 | 99.31 |
| LUMO+9 | 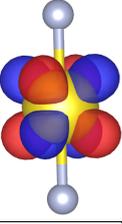 | 86.20 |
| LUMO+13 | 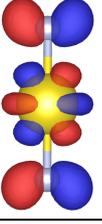 | 99.31 |
| LUMO+14 | 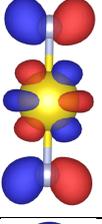 | 54.57 |
| LUMO+18 | 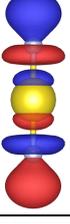 | 54.71 |

*PrNO*

| Orbital | Isosurface | % Pr $f$ |
|---|---|---|
| HOMO-2 | 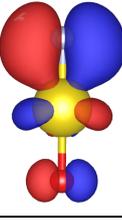 | 21.35 |
| HOMO-1 | 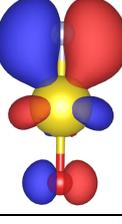 | 21.02 |
| HOMO | 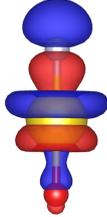 | 61.98 |
| LUMO | 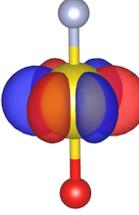 | 99.03 |
| LUMO+1 | 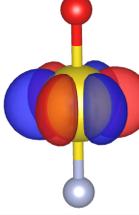 | 99.03 |
| LUMO+2 | 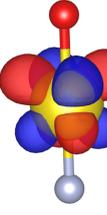 | 96.48 |
| LUMO+3 | 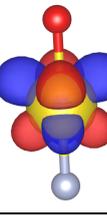 | 96.48 |
| LUMO+7 | 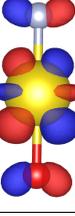 | 56.69 |
| LUMO+8 | 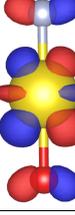 | 56.58 |
| LUMO+10 | 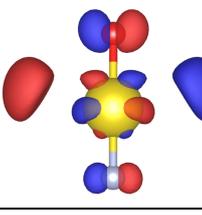 | 15.76 |
| LUMO+11 | 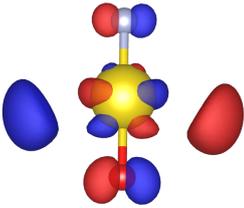 | 15.38 |

<div align="center"><em>PrNF⁺</em></div>

| Orbital | Isosurface | % Pr $f$ |
|---|---|---|
| HOMO-2 | 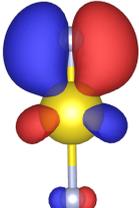 | 20.92 |
| HOMO-1 | 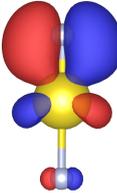 | 79.99 |
| HOMO | 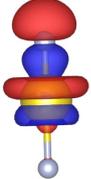 | 56.96 |
| LUMO | 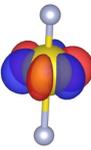 | 98.72 |
| LUMO+1 | 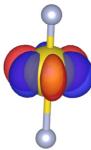 | 98.72 |
| LUMO+2 | 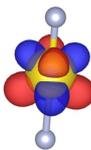 | 96.49 |

| Orbital | Isosurface | |
|---------|-----------|---|
| LUMO+3 | 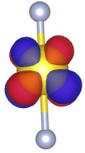 | 97.38 |
| LUMO+4 | 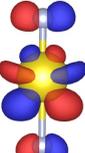 | 72.87 |
| LUMO+5 | 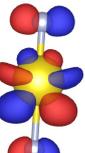 | 73.23 |
| LUMO+9 | 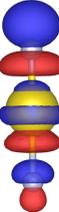 | 14.39 |

$PrNF_2$

| Orbital | Isosurface | % Pr $f$ |
|---------|-----------|----------|
| HOMO-2 | 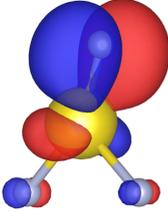 | 15.57 |
| HOMO-1 | 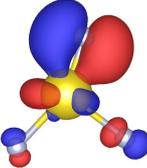 | 25.21 |
| HOMO | 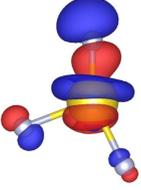 | 50.05 |
| LUMO | 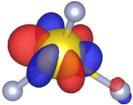 | 95.90 |
| LUMO+1 | 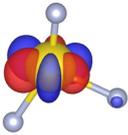 | 96.69 |
| LUMO+2 | 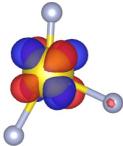 | 96.65 |
| LUMO+3 | 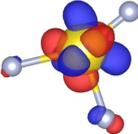 | 93.24 |
| LUMO+4 | 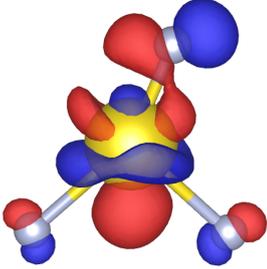 | 66.71 |
| LUMO+5 | 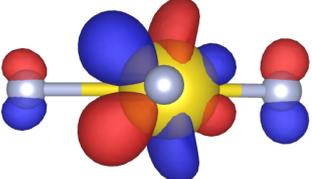 | 78.02 |
| LUMO+6 | 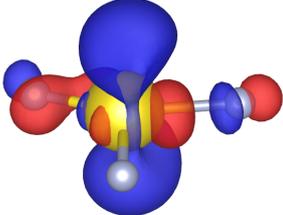 | 23.87 |

| LUMO+7 | 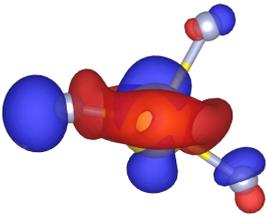 | 14.21 |

| Orbital | Isosurface | % Pr $f$ |
|---------|------------|----------|
| HOMO-2 | 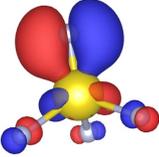 | 15.51 |
| HOMO-1 | 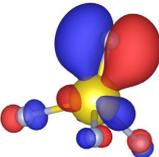 | 18.54 |
| HOMO | 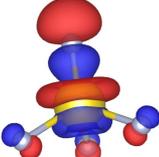 | 39.62 |
| LUMO | 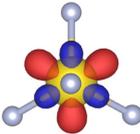 | 92.39 |
| LUMO+1 | 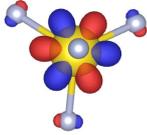 | 95.12 |
| LUMO+2 | 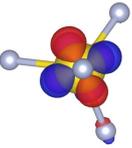 | 94.15 |
| LUMO+3 | 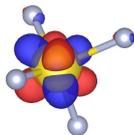 | 96.41 |
| LUMO+4 | 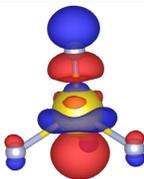 | 51.04 |
| LUMO+5 | 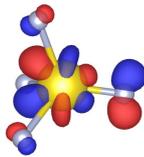 | 75.12 |
| LUMO+6 | 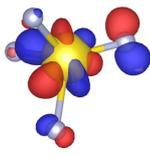 | 75.49 |

$PrO_2^+$

| Orbital | Isosurface | % Pr $f$ |
|---|---|---|
| HOMO-2 | 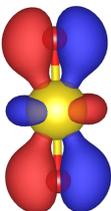 | 21.95 |
| HOMO-1 | 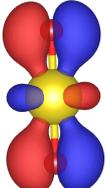 | 21.90 |
| HOMO | 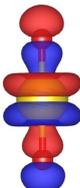 | 65.30 |
| LUMO | 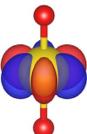 | 99.84 |
| LUMO+1 | 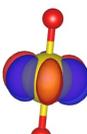 | 98.67 |
| LUMO+2 | 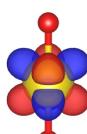 | 98.79 |
| LUMO+3 | 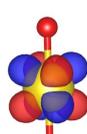 | 98.78 |
| LUMO+4 | 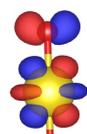 | 74.10 |
| LUMO+5 | 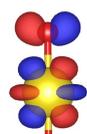 | 73.90 |
| | $PrOF_2^+$ | |
| Orbital | Isosurface | % Pr $f$ |
| HOMO-1 | 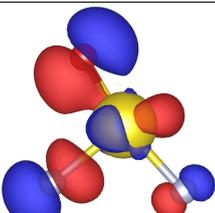 | 11.25 |

| HOMO | 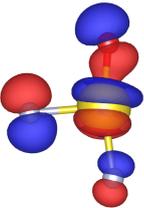 | 33.47 |
|---|---|---|
| LUMO | 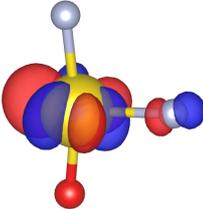 | 92.32 |
| LUMO+1 | 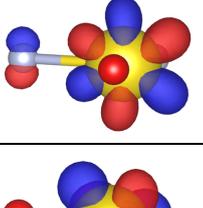 | 94.32 |
| LUMO+2 | 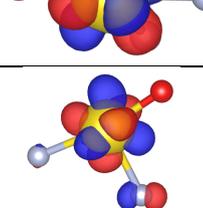 | 97.59 |
| LUMO+3 | 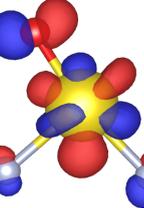 | 93.45 |
| LUMO+4 | 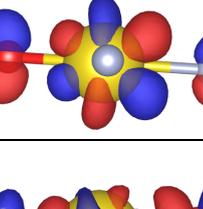 | 78.73 |
| LUMO+5 | 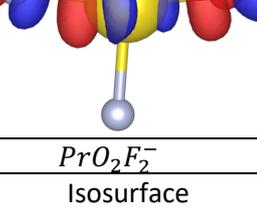 | 82.51 |
| LUMO+6 | 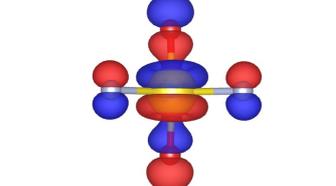 | 39.4 |

$$PrO_2F_2^-$$

| Orbital | Isosurface | % Pr $f$ |
|---|---|---|
| HOMO | 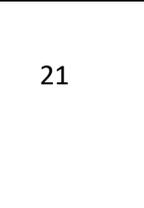 | 42.60 |
| LUMO | 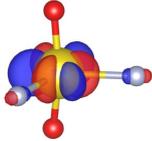 | 93.17 |
| LUMO+1 | 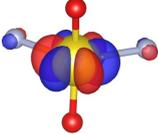 | 94.68 |
| LUMO+2 | 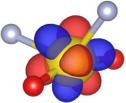 | 96.96 |
| LUMO+3 | 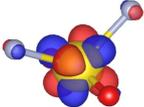 | 94.71 |
| LUMO+4 | 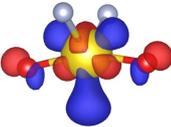 | 58.66 |
| LUMO+5 | 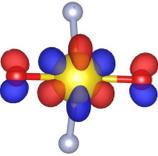 | 77.89 |
| LUMO+6 | 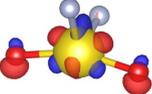 | 17.04 |
| LUMO+7 | 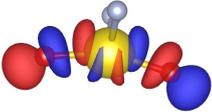 | 37.64 |

| PrOF₃ | | |
|---|---|---|
| Orbital | Isosurface | % Pr $f$ |
| HOMO | 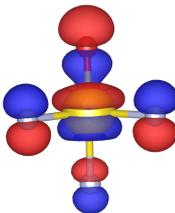 | 21.94 |
| LUMO | 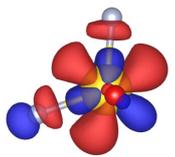 | 89.67 |
| LUMO+1 | 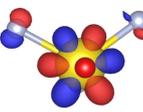 | 91.93 |
| LUMO+2 | 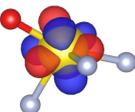 | 97.81 |
| LUMO+3 | 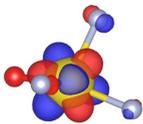 | 92.36 |
| LUMO+4 | 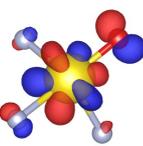 | 81.84 |
| LUMO+5 | 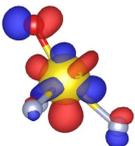 | 80,.49 |
| LUMO+6 | 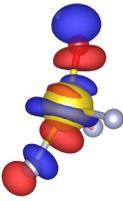 | 47.04 |

| PrF$_4$$^+$ | | |
|---|---|---|
| Orbital | Isosurface | % Pr $f$ |
| LUMO | 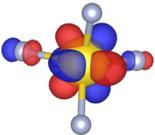 | 89.91 |
| LUMO+1 | 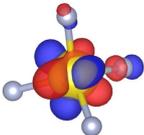 | 90.14 |
| LUMO+2 | 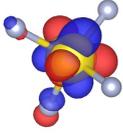 | 90.06 |
| LUMO+3 | 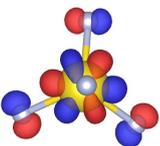 | 81.95 |
| LUMO+4 | 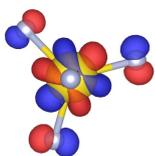 | 82.31 |
| LUMO+5 | 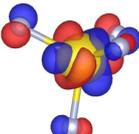 | 82.59 |
| LUMO+6 | 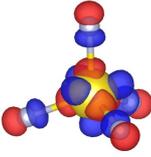 | 74.70 |

| PrF$_5$ | | |
|---|---|---|
| Orbital | Isosurface | % Pr $f$ |
| LUMO | 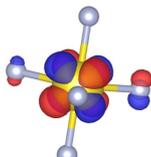 | 92.52 |
| LUMO+1 | 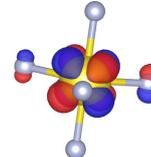 | 93.80 |
| LUMO+2 | 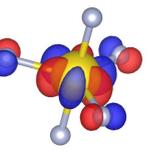 | 84.61 |
| LUMO+3 | 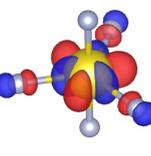 | 83.34 |
| LUMO+4 | 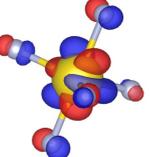 | 82.35 |
| LUMO+5 | 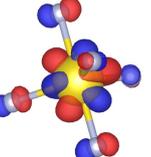 | 82.72 |

| LUMO+6 | 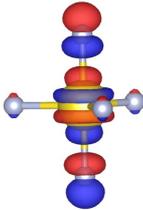 | 65.11 |

| Molecule | Multiplicity | Ionization energy [eV] | Electron affinity [eV] | Electronic Hardness [eV] |
|---|---|---|---|---|
| $PrN_2^-$ | Spinorbit | -0.5 | -2.8 | 2.3 |
| | Singlet | -0.5 | -2.9 | 2.4 |
| | Triplet | -0.5 | -2.8 | 2.3 |
| | Quintet | -1.0 | -1.0 | 0.1 |
| PrNO | Spinorbit | 6.0 | 2.7 | 3.3 |
| | Singlet | 6.1 | 2.4 | 3.7 |
| | Triplet | 6.6 | 2.6 | 4.0 |
| | Quintet | 5.2 | 2.5 | 2.7 |
| $PrNF^+$ | Spinorbit | 14.1 | 10.3 | 3.8 |
| | Singlet | 13.8 | 10.5 | 3.3 |
| | Triplet | 14.0 | 10.2 | 3.9 |
| | Quintet | 12.3 | 9.5 | 2.7 |
| $PrNF_2$ | Spinorbit | 7.7 | 4.5 | 3.2 |
| | Singlet | 7.6 | 4.1 | 3.4 |
| | Triplet | 7.7 | 4.5 | 3.2 |
| | Quintet | 6.2 | 3.3 | 2.9 |
| $PrNF_3^-$ | Spinorbit | 1.3 | -1.8 | 3.1 |
| | Singlet | 2.5 | -0.8 | 3.4 |
| | Triplet | 2.4 | -0.6 | 3.0 |
| | Quintet | 1.3 | -1.8 | 3.1 |
| $PrO_2^+$ | Spinorbit | 14.1 | 11.3 | 2.8 |
| | Singlet | 14.8 | 10.5 | 4.4 |
| | Triplet | 15.2 | 11.1 | 4.1 |
| | Quintet | 13.1 | 10.8 | 2.3 |
| $PrOF_2^+$ | Spinorbit | 16.1 | 12.2 | 3.9 |
| | Singlet | 16.1 | 12.0 | 4.1 |
| | Triplet | 16.2 | 12.8 | 3.4 |
| | Quintet | 13.0 | 11.3 | 1.7 |
| $PrO_2F_2^-$ | Spinorbit | 3.2 | -0.2 | 3.4 |
| | Singlet | 2.9 | -1.1 | 4.0 |
| | Triplet | 3.3 | -0.2 | 3.4 |
| | Quintet | 1.7 | -0.3 | 1.9 |
| $PrOF_3$ | Spinorbit | 9.8 | 6.0 | 3.8 |
| | Singlet | 9.9 | 5.9 | 4.0 |
| | Triplet | 10.0 | 6.9 | 3.1 |
| | Quintet | 7.3 | 6.0 | 1.3 |
| $PrF_4^+$ | Spinorbit | 17.8 | 14.3 | 3.5 |
| | Singlet | 17.9 | 14.2 | 3.7 |
| | Triplet | 16.7 | 13.2 | 3.5 |
| | Quintet | 14.1 | 12.4 | 1.7 |
| $PrF_5$ | Spinorbit | 11.5 | 7.6 | 3.9 |
| | Singlet | 11.6 | 7.5 | 4.1 |
| | Triplet | 11.4 | 7.1 | 4.3 |
| | Quintet | 7.8 | 6.7 | 1.1 |

| Equation | Calculations | ΔE [kcal/mol] | ΔH [kcal/mol] | ΔG [kcal/mol] |
|---|---|---|---|---|
| $PrN_2^- \rightarrow PrN^- + N$ | Scalar | 99.5 | 98.7 | 90.1 |
| $PrN_2^- \rightarrow PrN^- + N$ | Spinorbit | 94.2 | 93.7 | 86.7 |
| $PrN_2^- \rightarrow PrN + N^-$ | Scalar | 134.5 | 133.8 | 125.4 |
| $PrN_2^- \rightarrow PrN + N^-$ | Spinorbit | 114.2 | 113.9 | 107.1 |
| $PrN_2^- \rightarrow Pr^- + N_2$ | Scalar | 0.4 | 1.9 | -3.7 |
| $PrN_2^- \rightarrow Pr^- + N_2$ | Spinorbit | 14 | 15.8 | 11.8 |
| $PrN_2^- \rightarrow Pr + N_2^-$ | Scalar | 57.6 | 58.3 | 52.5 |
| $PrN_2^- \rightarrow Pr + N_2^-$ | Spinorbit | 67.1 | 68.1 | 63.9 |
| $2\,PrN_2^- \rightarrow 2\,PrN^- + N_2$ | Scalar | -27.4 | -26.3 | -36.4 |
| $2\,PrN_2^- \rightarrow 2\,PrN^- + N_2$ | Spinorbit | -38.0 | -36.3 | -43.3 |
| $PrNF^+ \rightarrow PrN + F^+$ | Scalar | 394.0 | 393.7 | 387.3 |
| $PrNF^+ \rightarrow PrN + F^+$ | Spinorbit | 360.3 | 360.0 | 350.4 |
| $PrNF^+ \rightarrow PrF^+ + N$ | Scalar | 41.1 | 40.3 | 33.5 |
| $PrNF^+ \rightarrow PrF^+ + N$ | Spinorbit | 45.3 | 44.6 | 34.5 |
| $PrNF^+ \rightarrow PrN^+ + F^\bullet$ | Scalar | 127.8 | 127.3 | 121 |
| $PrNF^+ \rightarrow PrN^+ + F^\bullet$ | Spinorbit | 94.8 | 94.2 | 84.5 |
| $2\,PrNF^+ \rightarrow 2\,PrN^+ + F_2$ | Scalar | 147.6 | 147.4 | 141.0 |
| $2\,PrNF^+ \rightarrow 2\,PrN^+ + F_2$ | Spinorbit | 147.7 | 147.1 | 134.2 |
| $2\,PrNF^+ \rightarrow 2\,PrF^+ + N_2$ | Scalar | -144.2 | -143.1 | -149.7 |
| $2\,PrNF^+ \rightarrow 2\,PrF^+ + N_2$ | Spinorbit | -135.8 | -134.7 | -147.6 |
| $2\,PrNF^+ \rightarrow PrN^+ + PrF^+ + NF$ | Scalar | 50.5 | 50.1 | 43.1 |
| $2\,PrNF^+ \rightarrow PrN^+ + PrF^+ + NF$ | Spinorbit | 54.9 | 54.4 | 41.0 |
| $PrNF_2 \rightarrow PrF + FN$ | Skalar | 113.6 | 112.9 | 104.6 |
| $PrNF_2 \rightarrow PrF + FN$ | Spinorbit | 117.6 | 117.3 | 109.6 |
| $PrNF_2 \rightarrow PrN + F_2$ | Skalar | 186.2 | 185.9 | 178.2 |
| $PrNF_2 \rightarrow PrN + F_2$ | Spinorbit | 191.3 | 191.3 | 184.2 |
| $PrNF_2 \rightarrow PrF_2^+ + N^-$ | Skalar | 216.4 | 215.5 | 208.1 |
| $PrNF_2 \rightarrow PrF_2^+ + N^-$ | Spinorbit | 200.8 | 200.2 | 193.4 |
| $PrNF_2 \rightarrow PrNF + F \bullet$ | Skalar | 133.1 | 132.4 | 124.4 |
| $PrNF_2 \rightarrow PrNF + F \bullet$ | Spinorbit | 99.1 | 98.7 | 91.4 |
| $2\,PrNF_2 \rightarrow 2\,PrNF + F_2$ | Skalar | 158.3 | 157.5 | 147.9 |
| $2\,PrNF_2 \rightarrow 2\,PrNF + F_2$ | Spinorbit | 156.4 | 156.2 | 148.0 |
| $PrNF_3^- \rightarrow PrF_2 + NF^-$ | Skalar | 132.1 | 130.5 | 119.6 |
| $PrNF_3^- \rightarrow PrF_2 + NF^-$ | Spinorbit | 143.5 | 142.8 | 134.0 |
| $PrNF_3^- \rightarrow PrF_2^- + NF$ | Skalar | 113.8 | 113.1 | 103.2 |
| $PrNF_3^- \rightarrow PrF_2^- + NF$ | Spinorbit | 128.4 | 127.7 | 118.9 |
| $PrNF_3^- \rightarrow PrNF + F_2^-$ | Skalar | 145.8 | 144.7 | 134.2 |
| $PrNF_3^- \rightarrow PrNF + F_2^-$ | Spinorbit | 154.8 | 154.2 | 145.0 |
| $PrNF_3^- \rightarrow PrNF^- + F_2$ | Skalar | 207.4 | 206.8 | 197.1 |
| $PrNF_3^- \rightarrow PrNF^- + F_2$ | Spinorbit | 220.3 | 220.2 | 211.6 |
| $PrNF_3^- \rightarrow PrNF_2^- + F \bullet$ | Skalar | 136.9 | 136.1 | 127.6 |
| $PrNF_3^- \rightarrow PrNF_2^- + F \bullet$ | Spinorbit | 118.5 | 118.3 | 110.5 |
| $2\,PrNF_3^- \rightarrow 2\,PrNF_2^- + F_2$ | Skalar | 165.8 | 165.0 | 154.2 |
| $2\,PrNF_3^- \rightarrow 2\,PrNF_2^- + F_2$ | Spinorbit | 195.1 | 195.3 | 186.1 |